\documentclass[11pt,a4paper]{article}
\usepackage{jcappub}
\usepackage{amsmath,graphicx,hyperref,bm,multirow}
\begin{document}

	\title{The observer-dependent shadow of the Kerr black hole}
	
\author{Zhe Chang,}
\author[1]{and Qing-Hua Zhu,\note{Corresponding author.}}

\affiliation{Institute of High Energy Physics, Chinese Academy of Sciences, Beijing 100049, China}
\affiliation{University of Chinese Academy of Sciences, Beijing 100049, China}

\emailAdd{zhuqh@ihep.ac.cn}

\arxivnumber{2104.14221}

\abstract{
Motivated by inclination of the Earth's orbit that is not located at galactic plane for observing the shadow of Sgr A*, we consider the black hole shadow for arbitrary inclinations and different velocities of observers. A surprising finding of the study is that rotation axis of a black hole might not be extracted from its shadow, since the ways of the shadow getting distorted depend not only on the spin of the black hole, but also velocities of observers. Namely, appearance of the shadow could be rotated by an angle in observers' celestial sphere due to the observer in motion. In order to further confirm this result, a formalism is presented for calculating the shadow in terms of the velocity perturbations.  Besides, we also consider the Earth's orbit for the shadow of Sgr A* by making use of this formalism. 
}
	
{\maketitle}

\section{introduction}

Since the first image of the black hole taken by Event Horizon Telescopes (EHT) in 2019 \cite{collaboration_first_2019,collaboration_first_2019-1}, the observation on the shadow of black hole is promising to be a direct way to test general relativity in the strong field regime \cite{moffat_black_2015,moffat_modified_2015,moffat_masses_2020,kumar_black_2020,jusufi_rotating_2020,zhu_shadows_2019,tian_testing_2019,ovgun_shadow_2018,hennigar_shadows_2018,ayzenberg_black_2018,konoplya_general_2016,li_measuring_2014,wei_observing_2013,johannsen_photon_2013,atamurotov_shadow_2013,amarilla_null_2010}. Based on the seed studies of Synge \cite{synge_escape_1966} and Bardeen \cite{bardeen_timelike_1973-1}, the shadow can be a probe of a black hole in the present of electromagnetism \cite{allahyari_magnetically_2020,kumar_shadow_2019,wang_shadows_2018,chael_role_2018,gold_probing_2017}, accretion \cite{reji_gravitational_2020,haroon_shadow_2020,shaikh_shadows_2019,bambi_testing_2019,taylor_exploring_2018,perlick_light_2017,akiyama_imaging_2017,perlick_influence_2015,atamurotov_optical_2015}, dark matter \cite{jusufi_black_2019,jusufi_shadows_2020,konoplya_shadow_2019,haroon_shadow_2019,cunha_shadows_2016}, exotic matter \cite{roy_evolution_2020,davoudiasl_ultralight_2019,cunha_lensing_2017,abdujabbarov_shadow_2017,vincent_imaging_2016,vincent_astrophysical_2016,ohgami_wormhole_2015,nedkova_shadow_2013} , or even quantum effect \cite{liu_shadow_2020,hu_qed_2020,giddings_event_2018,wang_shadow_2017,amarilla_shadow_2012}.

In order to extract information of a black hole via its shadow, one should notice that the appearance of the shadow is not determined by the black hole alone. There are also influences on the shadow from inclination angle between the observer's line-of-sight and the spin axis of the black hole \cite{wang_shadow_2020,li_shadow_2020,liu_shadow_2020,banerjee_silhouette_2020,wei_curvature_2019,tsukamoto_black_2018,ovgun_shadow_2018,tsupko_analytical_2017,amir_shapes_2016,papnoi_shadow_2014,nedkova_shadow_2013,johannsen_photon_2013,johannsen_testing_2010,takahashi_shapes_2004}, or observers' comoving with the expansion of the Universe \cite{vagnozzi_concerns_2020,tsupko_black_2020,tsupko_first_2019,firouzjaee_black_2019,stuchlik_light_2018,perlick_black_2018,eiroa_shadow_2018}. Namely, the appearance of black hole shadow is observer-dependent in principle. It is non-trivial study on black hole shadow in the view of observers located at curved space-time \cite{perlick_black_2018,grenzebach_photon_2014,chang_revisiting_2020,chang_does_2020}. For example, the comoving observers would view a finite size of the shadow even when they go through the cosmological horizon and approach the infinity \cite{perlick_black_2018,chang_black_2020}. For an observer orbiting a Kerr black hole at its rotation plane, the shape of the shadow could tend to be circular \cite{chang_does_2020}.

In recent studies on black hole shadow, a new approach for calculating the shadow was proposed \cite{chang_revisiting_2020,chang_does_2020} and used in recent study \cite{He:2020dfo}. This formalism is suited for studying the shadow with respect to observers in motion. In this paper, we will investigate influence of Earth's orbit on the shadow of Sagittarius A* (Sgr A*) based on this approach. At the first step, we extend the previous studies \cite{chang_revisiting_2020,chang_does_2020} to arbitrary inclination, and observers at different velocities. It is partly motivated by the motion of the Earth with respect to the Sgr A*, where the inclination of the Earth's orbit are not located at galactic plane. It seems to be common sense that distorted shape of the shadow indicates rotation axis of a black hole \cite{bardeen_timelike_1973-1}. However, in this study, it is found that the appearance of the shadow would be rotated by a certain angle in celestial sphere in the view of observers moving towards direction of changing inclination $\theta$. 
Second, in order to handle local orbits of observers with respect to a black hole, we present a formalism for calculating the shadow in terms of local velocity expansion. 
It shows that influence of the orbital velocity of the Earth on the shadows of Sgr A* is much larger than that of the displacement in Earth's orbit.

It is worth mentioning that the numerical studies on appearance of the black hole shadow from a virtual reality journey also investigated 
the influence from observers in motion  \cite{james_gravitational_2015,davelaar_observing_2018}. In the previous references, orthonormal tetrads have been used as a local frame of observers. Alternatively, in this paper, we use the approach developed recently \cite{chang_revisiting_2020,chang_does_2020}, which can extract information of observers without using the orthonormal tetrads. More theoretical comparisons of these two approaches were also presented in these Refs \cite{chang_revisiting_2020,chang_does_2020}. In principle, the latter approach is compatible with the studies on numerical or parameterized black hole from its shadow \cite{moscibrodzka_observational_2014,chan_fast_2015,mizuno_current_2018,davelaar_observing_2018,younsi_new_2016},

The rest of the paper is organized as follows. In section~\ref{II}, we brief review the astrometric observable approach for calculating the black hole shadow. The formula of distortion parameter is updated for observers in arbitrary motion. In section~\ref{III}, we present the results of the shadow influenced by observers at different velocities. The shadows with respect to an observer moving towards direction of changing inclination $\theta$ is showed to be very interesting. In section~\ref{IV}, we introduce a formalism of black hole shadow in the present of local velocity perturbations. And we will present the order of magnitude estimation of the shadow affected  the Earth's orbit. Finally, conclusions and discussions are given in Section~\ref{V}.

\section{Brief review of astrometric observable approach for the shadow of rotating black hole}\label{II}

For a general rotating black hole, it can be parametrized as
\begin{equation}
  {\rm{d}} s^2 = - N^2 ({\rm{d}} t + A {\rm{d}} \phi)^2 + G_r^2 {\rm{d}} r^2 +
  G_{\theta}^2 {\rm{d}} \theta^2 + G_{\phi}^2 {\rm{d}} \phi^2 ~ .
\end{equation}
In order to utilize analytic formulae for evaluating the shadows, we consider
an asymmetric and stationary space-time equipped with the Carter's constant. For
Kerr black hole in the Boyer-Lindquist coordinate, we have
\begin{subequations}
\begin{eqnarray}
  N^2 & = & - g_{t   t} = 1 - \frac{2 M   r}{\Sigma} ~, \\
  A & = & \frac{g_{t \phi}}{g_{t   t}} = \frac{2 a   r \sin^2
  \theta}{\Sigma + 2 M   r} ~, \\
  G_r^2 & = & g_{r   r} = \frac{\Sigma}{\Delta} ~, \\
  G_{\theta}^2 & = & g_{\theta \theta} = \Sigma ~, \\
  G_{\phi}^3 & = & g_{\phi \phi} - \frac{(g_{t \phi})^2}{g_{t   t}} =
  \left( r^2 + a^2 + \frac{2 M   r a^2 \sin^2 \theta}{\Sigma - 2 M
    r} \right) \sin^2 \theta ~, 
\end{eqnarray}
\end{subequations}
where $a$ is the spin parameter of Kerr black hole and $\Sigma = r^2 + a^2
\cos^2 \theta$.

\subsection{Hamilton-Jacobi method for geodesics and Photon sphere}

By making use of Hamilton-Jacobi equation for geodesic in the space-time of
rotating black hole, we have
\begin{eqnarray}
  \epsilon & = & - \frac{1}{N^2} \left( \frac{\partial \mathcal{S}}{\partial
  t} \right)^2 + \frac{1}{G_r^2} \left( \frac{\partial \mathcal{S}}{\partial
  r} \right)^2 + \frac{1}{G_{\theta}^2} \left( \frac{\partial
  \mathcal{S}}{\partial \theta} \right)^2 + \frac{1}{G_{\phi}^2} \left( A
  \frac{\partial \mathcal{S}}{\partial t} - \frac{\partial
  \mathcal{S}}{\partial \phi} \right)^2 ~,  \label{7}
\end{eqnarray}
where $p_{\mu} = \frac{\partial \mathcal{S}}{\partial x^{\mu}}$. Since the
space-time is assumed to be asymmetric and stationary, there are two integral
constants $E$ and $L$ along directions of increasing $t$ and $\phi$, namely
\begin{subequations}
\begin{eqnarray}
  p_t & = & - E ~,  \label{8}\\
  p_{\phi} & = & L ~ .  \label{9}
\end{eqnarray} \label{A4}
\end{subequations}
Therefore, based on Eqs.~(\ref{7}) and (\ref{A4}), we have a
solution of $\mathcal{S}$ from complete integral, namely
\begin{eqnarray}
  \mathcal{S} & = & - E   t + S (r) + S (\theta) + L \phi ~ . 
\end{eqnarray}
The Eq.~(\ref{7}) can also be rewritten as
\begin{eqnarray}
  \epsilon & = & - \left( \frac{E}{N} \right)^2 + \left( \frac{p_r}{G_r}
  \right)^2 + \left( \frac{p_{\theta}}{G_{\theta}} \right)^2 + \left( \frac{A
    E + L}{G_{\phi}} \right)^2 ~ .  \label{11}
\end{eqnarray}
We consider space-time of rotating black hole in the present of Carter's
constant $K$. This indicates that we can separate variations $r, \theta$ by
multiplying a function $B (r, \theta)$ in the Eq.~(\ref{11}),
\begin{eqnarray}
  \left( \frac{B}{G_r} p_r \right)^2 + \left( \frac{B}{G_{\theta}} p_{\theta}
  \right)^2 & = & \epsilon B^2 + E^2 \left( \frac{B  }{N} \right)^2 -
  E^2 \left( \frac{B}{G_{\phi}} \right)^2 (A + \lambda)^2 + K - K ~ , \label{A7}
\end{eqnarray}
where we have defined $\lambda  \equiv  \frac{L}{E}$. Thus, the above equations can be rearranged as
\begin{subequations}
\begin{eqnarray}
  \Delta_r (r) p_r^2 & = & E^2 (f_1 (r ; \lambda) - \kappa) + \epsilon B^2_r
  ~,  \label{12}\\
  \Delta_{\theta} (\theta) p_{\theta}^2 & = & E^2 (\kappa + f_2 (\theta ;
  \lambda)) + \epsilon B_{\theta}^2 ~,  \label{13}
\end{eqnarray} \label{A9}
\end{subequations}
where we have used the notations,
\begin{eqnarray}
  \kappa & \equiv & \frac{K}{E^2} ~, \\
  \Delta_r (r) & \equiv & \left( \frac{B}{G_r} \right)^2 ~, \\
  \Delta_{\theta} (\theta) & \equiv & \left( \frac{B}{G_{\theta}} \right)^2
  ~, 
\end{eqnarray}
and
\begin{eqnarray}
  B_r^2 (r) + B^2_{\theta} (\theta) & \equiv & B^2 (r, \theta) ~, \\
  f_1 (r ; \lambda) + f_2 (\theta ; \lambda) & \equiv & \left( \frac{B
   }{N} \right)^2 - \left( \frac{B}{G_{\phi}} \right)^2 (A + \lambda)^2
  ~ . 
\end{eqnarray}
Based on Eqs.~(\ref{A4}),(\ref{A7}) and (\ref{A9}), 4-velocities
of geodesic can be summaries as
\begin{subequations}
\begin{eqnarray}
		p^t & = & E \left( \frac{1}{N^2} - \frac{A}{G_{\phi}^2} (A   +
		\lambda) \right) ~,  \label{20}\\
		p^r & = & \sigma_r E \frac{\sqrt{\Delta_r}}{B^2} \sqrt{f_1 (r ; \lambda) -
			\kappa + \epsilon \left( \frac{B_r}{E} \right)^2} ~, \\
		p^{\theta} & = & \sigma_{\theta} E \frac{\sqrt{\Delta_{\theta}}}{B^2}
		\sqrt{\kappa + f_2 (\theta ; \lambda) + \epsilon \left( \frac{B_{\theta}}{E}
			\right)^2} ~, \\
		p^{\phi} & = & \frac{E}{G_{\phi}^2} (A + \lambda) ~,  \label{23}
\end{eqnarray}\label{A16}
\end{subequations}
where $\sigma_r, \sigma_{\theta} = \pm$, since $p^r$ and $p^{\theta}$ could be
positive or negative. For time-like test particles, we have $\epsilon = - 1$.
And for null test particles, we have $\epsilon = 0$.

\begin{figure}[h]
	\centering
  \includegraphics[width=\linewidth]{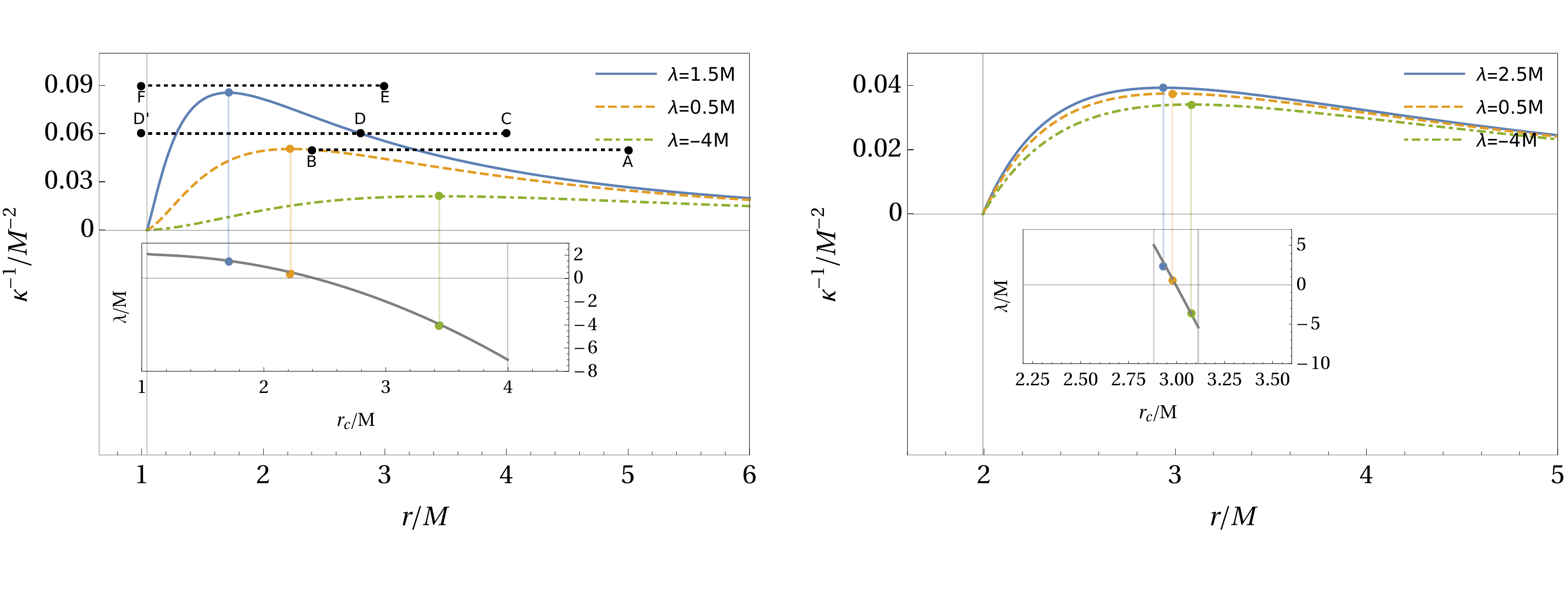}
  \caption{The $\kappa, \lambda$ as functions of $r_c$ in the space-time of
  Kerr black hole. We select spin parameter $a = 0.999M$ and $a = 0.1M$ in left
  panel and right panel, respectively.  \label{F1}}
\end{figure}

For studies on the black hole shadow, we consider out-going light rays, namely
$\sigma_r = 1$ and $\epsilon = 0$. For the light rays described in
Eqs.~(\ref{A16}), the photon sphere determines which types of the light rays from outside of black
hole could approach the black hole and then escape again. This critical
condition is the same as unstable circular orbits of light rays, namely
\begin{subequations}
	\begin{eqnarray}
		\left( \frac{{\rm{d}} r}{{\rm{d}} \lambda} \right)_{r_c} & = & 0 ~, \\
		\left( \frac{{\rm{d}}^2 r}{{\rm{d}} \lambda^2} \right)_{r_c} & = & 0 ~ . 
	\end{eqnarray}
\end{subequations}
It leads to
\begin{subequations}
	\begin{eqnarray}
		f_1 (r_c ; \lambda) - \kappa & = & 0 ~,  \label{26}\\
		\frac{{\rm{d}}}{{\rm{d}} r} f_1 (r_c ; \lambda) & = & 0 ~ .  \label{27}
	\end{eqnarray} \label{A-16}
\end{subequations}
Therefore, we can express $\lambda, \kappa$ in terms of $r_c$ by solving Eqs.~(\ref{A-16}). Due to
$(p^{\theta})^2 \geqslant 0$ and the expression of $\kappa(r_c)$, we can obtain the range of $r_c$ via
\begin{eqnarray}
  f_1 (r_c ; \lambda (r_c)) + f_2 (\theta ; \lambda (r_c)) & \geqslant & 0
  ~ . 
\end{eqnarray}
It indicates that the values of $r_c$ are different for observers at different
inclinations $\theta  $.
In left panel of Figure~\ref{F1}, the $\kappa, \lambda$ as function of $r_c$ are shown for Kerr black hole in detail. A light
ray with  integral constants $\lambda = 1.5M$ and $\kappa^{- 1} > 0.09M^2$ must
fall into the black hole, which is shown in straight line $E \rightarrow F$ in the plots.
Light rays with the same integral constants $\kappa^{- 1}$ but different values of
$\lambda$ could have different endings. For a light ray with $\kappa =
0.06M^2$ and $\lambda = 1.5M$, it propagates along straight line in schematic
diagram $C \rightarrow D \rightarrow C \rightarrow \infty$. For a light ray with $\kappa = 0.06M^2$ and $\lambda = 0.5M$, it would fall into the black
hole, namely the straight line \ $C \rightarrow D \rightarrow D'$. In the same
way, we can read integral constants of the light rays denoted by straight
line $A \rightarrow B \rightarrow A \rightarrow \infty$. It is a light rays with $\kappa = 0.05M^2$ and
$\lambda = 0.5M$. Whether a light ray can escape from the black hole is
determined by both the integral constants $\kappa$ and $\lambda$. In right panel of Figure~\ref{F1},
we present the $\kappa, \lambda$ as functions of $r_c$ for different spin
parameters $a$. It is not surprising that the range of $r_c$ tend to be around
$3 M$ as $a \rightarrow 0$.

\subsection{Measurement and quantification of the shadow of black holes}

The boundary of the shadow in observers' celestial sphere are determined by
the light rays from unstable circular orbits. To be specific, we can consider
light rays $k, w$ and $l$ from the unstable circular orbits \cite{chang_revisiting_2020}. The $k$ and
$w$ denote the light rays from the photon sphere at $r_{c,
\min}$ and $r_{c, \max}$, namely
\begin{subequations}
	\begin{eqnarray}
		k & = & p|_{r_c = r_{c, \min}} ~, \\
		w & = & p|_{r_c = r_{c, \max}} ~, 
	\end{eqnarray}
\end{subequations}
where 4-velocity $p$ is shown in Eqs.~(\ref{A16}). And the expression of
light ray $l$ is simply
\begin{eqnarray}
  l & = & p|_{r_c} ~ .  \label{31}
\end{eqnarray}
For observers located at rotation plane of a rotating black hole, the light rays $k$ and $w$ propagate within the rotation plane, since $\theta$-components of the 4-velocities $k$ and $w$ remain vanished. 
The appearance of the shadow can be obtained via accident angles between $l$ and $w$, $l$ and $k$, and $k$ and $w$. These angles are formulated as
\begin{subequations}
	\begin{eqnarray}
		\cos \gamma & = & \frac{k \cdot w}{(u \cdot k) (u \cdot w)} + 1 ~,
		\label{32}\\
		\cos \alpha & = & \frac{k \cdot l}{(u \cdot k) (u \cdot l)} + 1 ~, \\
		\cos \beta & = & \frac{l \cdot w}{(u \cdot l) (u \cdot w)} + 1 ~ . 
		\label{34}
	\end{eqnarray}\label{E10}
\end{subequations}
The $\alpha$, $\beta$ and $\gamma$ depend on the light rays from photon sphere
and 4-velocity of an observer $u$.

By making use of spherical trigonometric, we can express celestial coordinate
$\Psi$ and $\Phi$ in terms of the $\alpha, \beta$ and $\gamma$ , i.e.
\begin{subequations}
	\begin{eqnarray}
		\Psi & = & \arccos \left( \sigma_{\theta} \sin \beta \sqrt{1 - \left(
			\frac{\cos \alpha - \cos \beta \cos \gamma}{\sin \beta \sin \gamma}
			\right)^2} \right) ~,  \label{35}\\
		\Phi & = & \gamma - \arccos \left( \frac{\cos \beta}{\sin \Psi} \right) ~
		.  \label{36}
	\end{eqnarray} \label{A21}
\end{subequations}
Therefore, one can sketch the appearance of the shadow via the curve $(\Phi
(r_c), \Psi (r_c))  $ in observers' celestial sphere.

\begin{figure}[h]
	\centering
  \includegraphics[width=.4\linewidth]{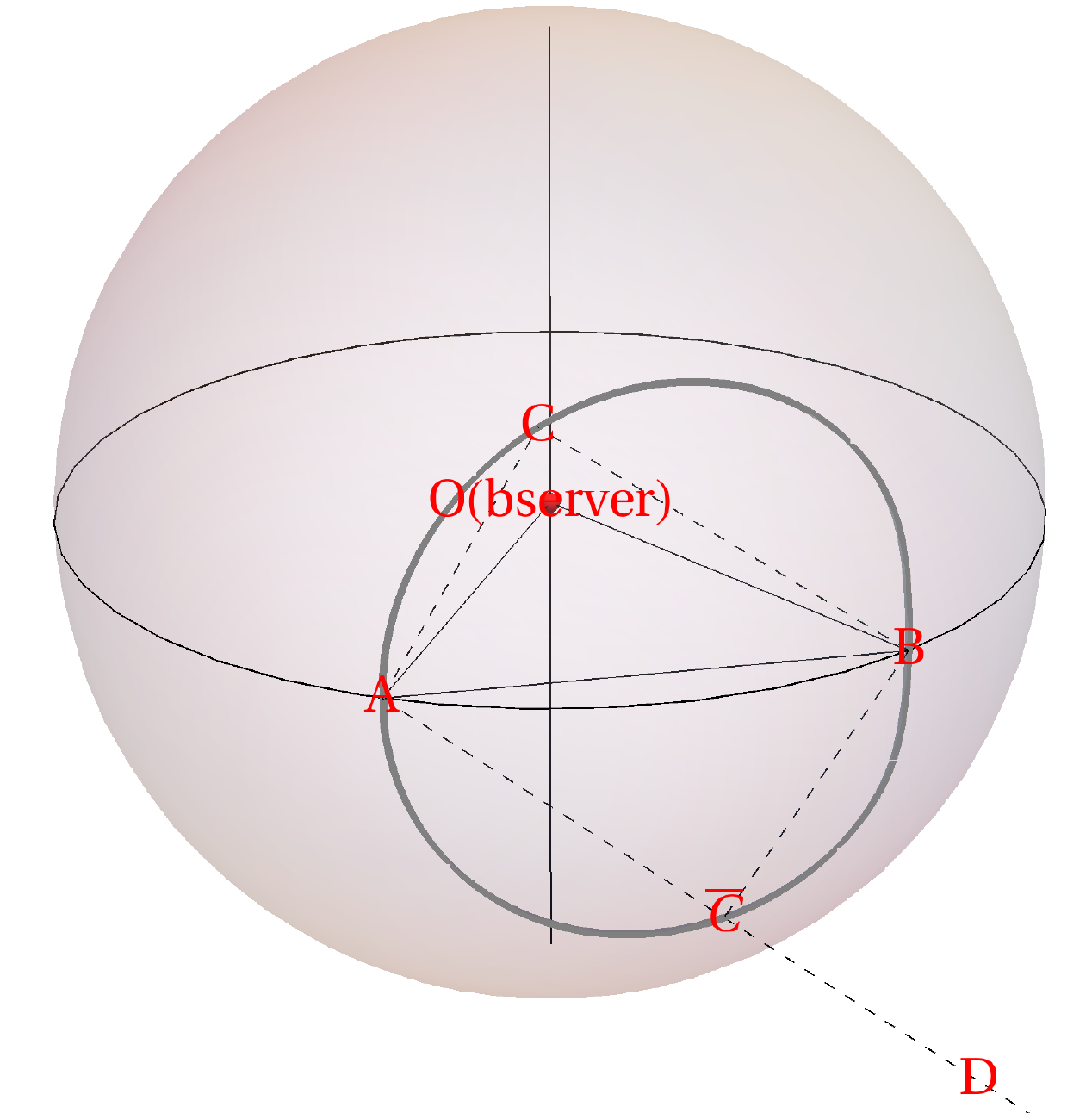}
  \caption{Schematic diagram of the distortion parameter $\cos \Xi$ in observers celestial sphere.\label{F2}}
\end{figure}

However, \ merely the sketch of the shadow is
not enough. In order to quantify the shape of the shadow, different distortion
parameters have been introduced
{\cite{hioki_measurement_2009,wei_curvature_2019,wei_intrinsic_2019,abdujabbarov_coordinate-independent_2015,tsupko_analytical_2017}}.
Here, by making use of the angles in Eqs.~(\ref{E10}), we can define
a distortion parameter $\cos\Xi$ in the form of {\cite{chang_does_2020}},
\begin{eqnarray}
  \cos \Xi & \equiv & \frac{1 + \cos \gamma - \cos \alpha - \cos \beta}{2
  \sigma_{\theta} \sqrt{(1 - \cos \alpha) (1 - \cos \beta)}} ~,  \label{37}
\end{eqnarray}
For $\cos \Xi \equiv 0$, the shape of the shadow is circular. And the previous
study {\cite{chang_does_2020}} suggests that the $\cos \Xi \neq 0$ indicates
the distortion of the shape of the shadow from circularity. The schematic
diagram is shown in Figure~\ref{F2}. The angle $\Xi$ is defined as
\begin{equation}
  \Xi \equiv \angle \rm{ACB} = \angle B \bar{C} D ~.
\end{equation}
However, we will show in section~\ref{subsection} that it is not true in a general case.
We can further find that $\cos \Xi = \rm{const.} $ also indicates shape of the shadow being circular.

Rearranging Eqs.~(\ref{A21}) and (\ref{37}), we obtain
equation in celestial coordinate $(\Phi, \Psi)$ for the boundary of the black
hole shadow, namely,
\begin{eqnarray}
  0 & = & (1 + \cos \gamma - \sin \Psi \cos \Phi - \sin \Psi \cos (\gamma - \Phi))^2 \nonumber\\ 
  & & -  4 \cos^2 \Xi (1 - \sin \Psi \cos \Phi) (1 - \sin \Psi \cos (\gamma - \Phi)) ~.
\end{eqnarray}
Here the $\gamma$ has no relevance with $\Phi$ and $\Psi$, while the $\Xi$
depends on the $\Phi$ or $\Psi$. The information about the appearance of black
hole shadow is included in the $\gamma$ and distortion parameter $\cos \Xi$.
The $\gamma$ describes characteristic size of the shadow, and the angle $\Xi$
describes the shape of the shadow. Therefore, in the following, we would focus
on the quantities $\Xi$ and $\gamma$ influenced by inclination and
motion of observers.

Besides, for the sake of intuitive, the shadow can be also sketched in a
2D-projection plane. Namely, one can transform the celestial spherical
coordinates into {\cite{chang_revisiting_2020}},
\begin{subequations}
	\begin{eqnarray}
		Y & \equiv & \frac{2 \sin \Phi \sin \Psi}{1 + \cos \Phi \sin \Psi} ~, \\
		Z & \equiv & \frac{2 \cos \Psi}{1 + \cos \Phi \sin \Psi} ~ . 
	\end{eqnarray}
\end{subequations}

\section{Influence of inclination and motion of observers on black hole
shadow}\label{III}

It might be the most interesting case that the black hole shadow are in the
view of the observers located at rotation plane of a rotating black
hole. And in this case, the shape of the shadow get the most distorted. However, it is usually not a realistic situation. The influence of inclination of
black hole on the shadow also deserves to be studied {\cite{tian_testing_2019,wang_shadow_2020,li_shadow_2020,liu_shadow_2020,banerjee_silhouette_2020,wei_curvature_2019,tsukamoto_black_2018,ovgun_shadow_2018,tsupko_analytical_2017,amir_shapes_2016,papnoi_shadow_2014,nedkova_shadow_2013,johannsen_photon_2013,johannsen_testing_2010,takahashi_shapes_2004}}. We
will further show that it is non-trivial, especially for the shadow in the view
of observers at finite distance. For illustration, we consider Kerr black hole
for example.

First, we consider black hole shadow in the view of static observers
for different inclination of rotating black hole. Next, we will turn to
the situation of geodesic observers with different velocities. For
observers moving along the direction of changing $\theta$, the appearance of the
shadow shows to be very interesting.

\subsection{Static observers}

\begin{figure}[h]
	\centering
  \raisebox{0.0\height}{\includegraphics[width=1\linewidth]{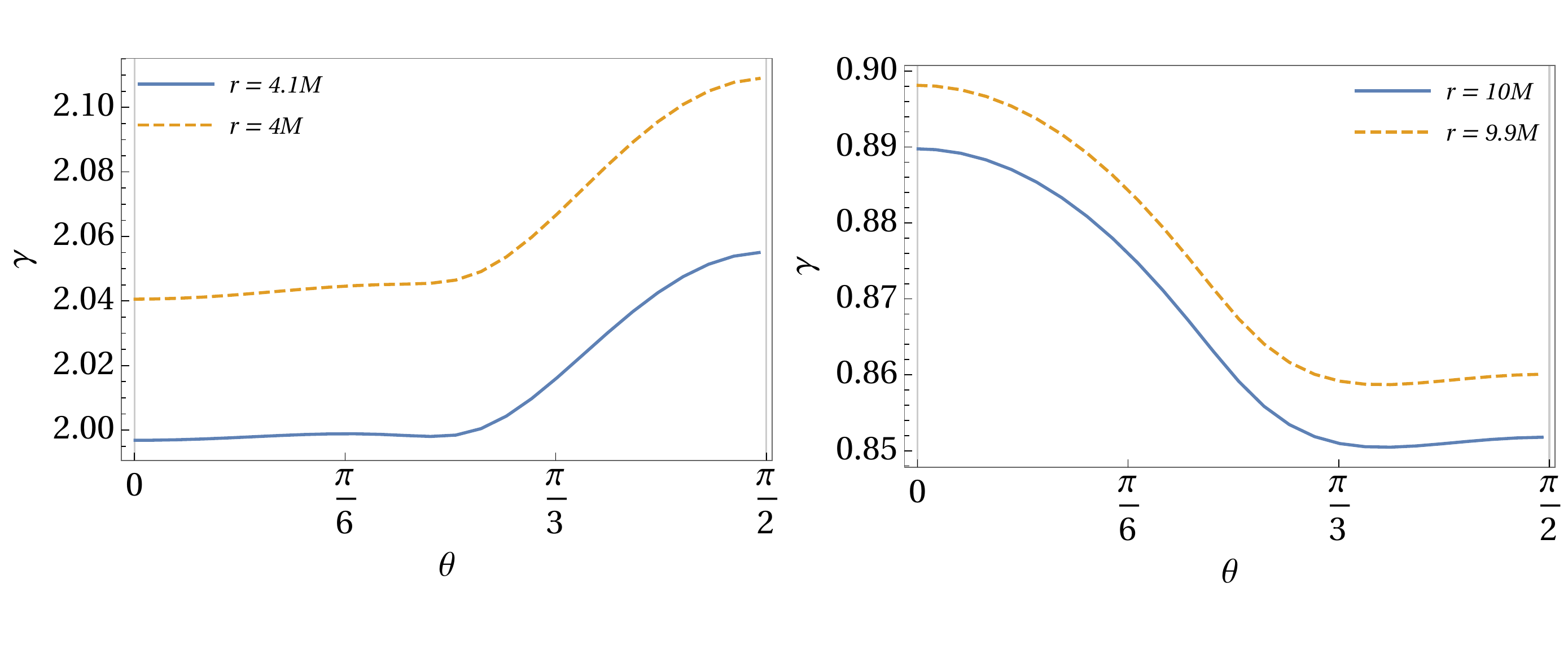}}
  \caption{Characteristic size of the shadow $\gamma$ as function of
  inclination $\theta$ for selected distance $r$. Here, the spin parameter $a$ is set to be
  $0.999M$.\label{F3}}
\end{figure}

\begin{figure}[h]
	\centering
	\raisebox{0.0\height}{\includegraphics[width=1\linewidth]{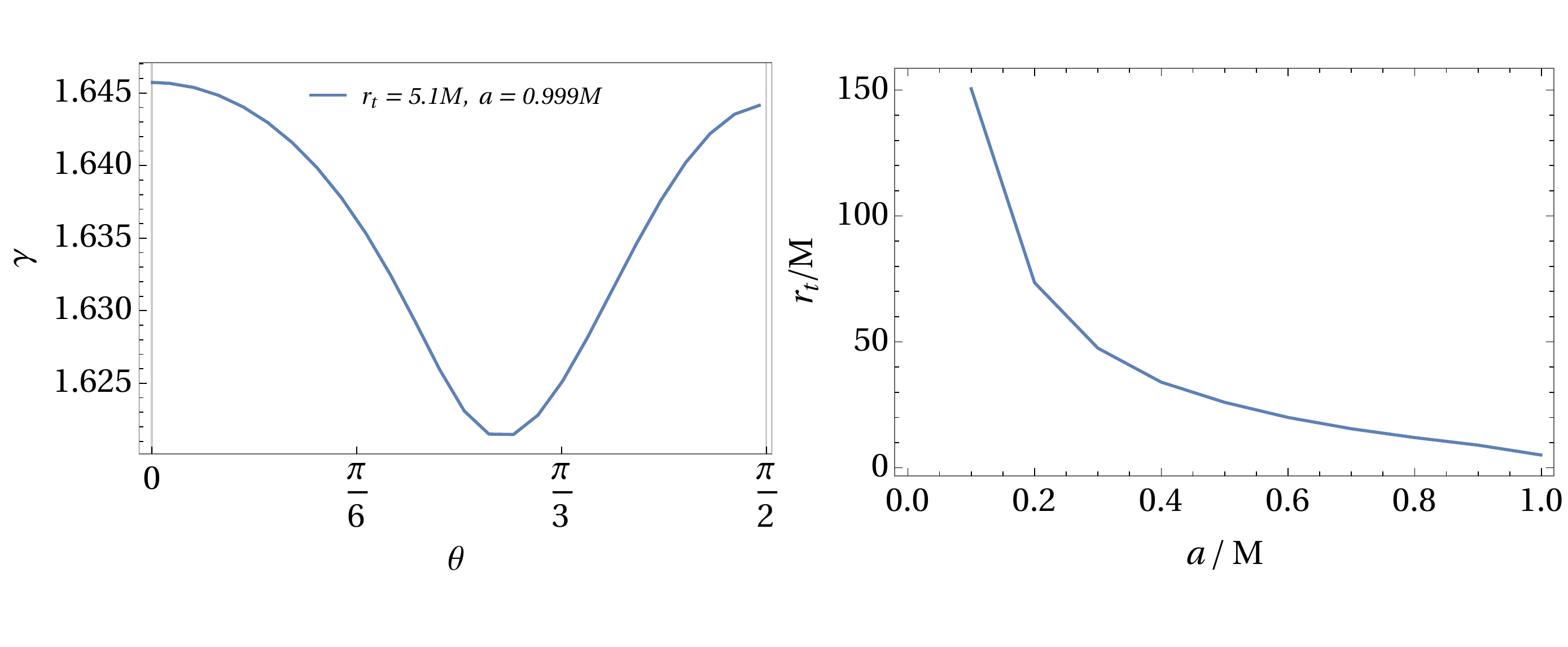}}
	\caption{Left panel: Characteristic size of the shadow $\gamma$ as function of inclination $\theta$ for the transition distance $r_{\rm{t}}$. Right panel: Transition distance $r_{\rm{t}}$ as function of spin parameter $a$.\label{F3A}}
\end{figure}

\begin{figure}[h]
	\centering
  \raisebox{0.0\height}{\includegraphics[width=16.31031090122cm,height=6.6690935327299cm]{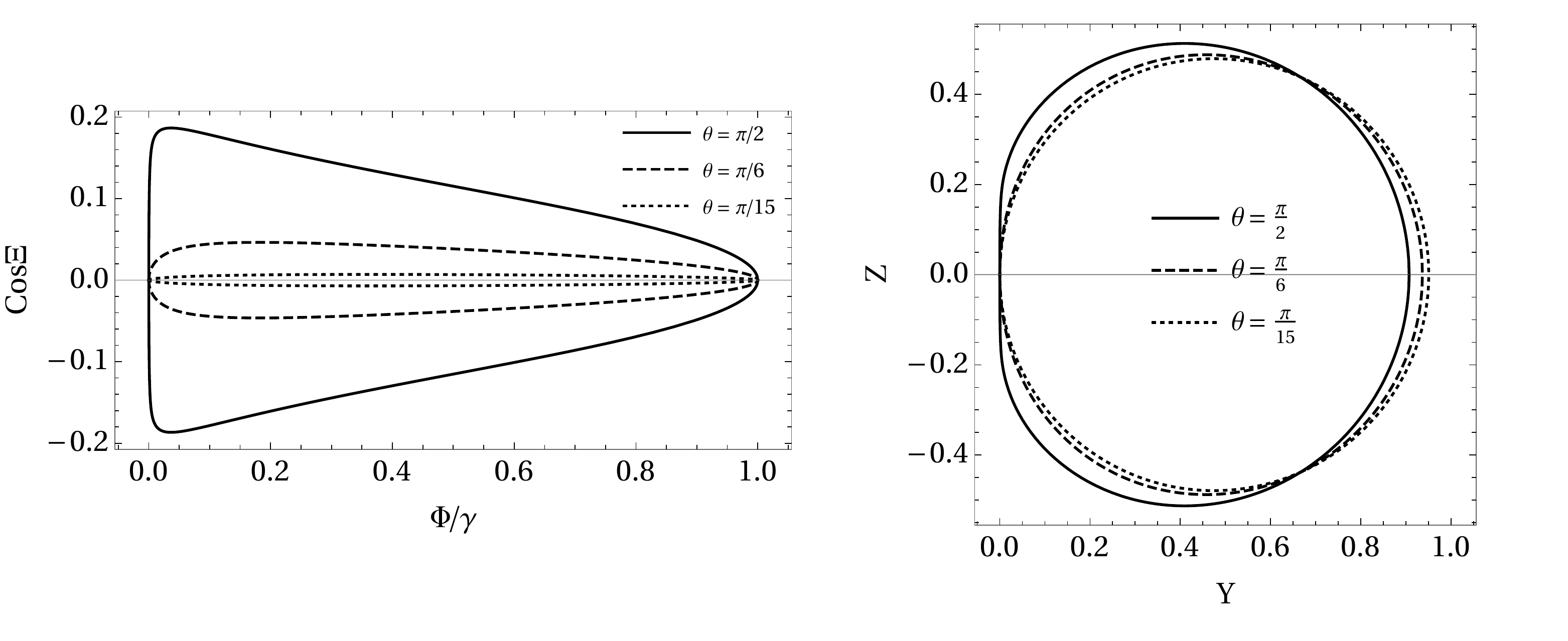}}
  \caption{Left panel: distortion parameter $\cos \Xi  $ as function of
  $\Phi / \gamma$ for different inclination $\theta$. Right panel: the
  appearance of the shadow in projection plane for different inclination.
  Here, we let observers' position $r = 10 M$.\label{F4}}
\end{figure}

\begin{figure}[h]
	\centering
	\raisebox{0.0\height}{\includegraphics[width=16.31031090122cm,height=6.6690935327299cm]{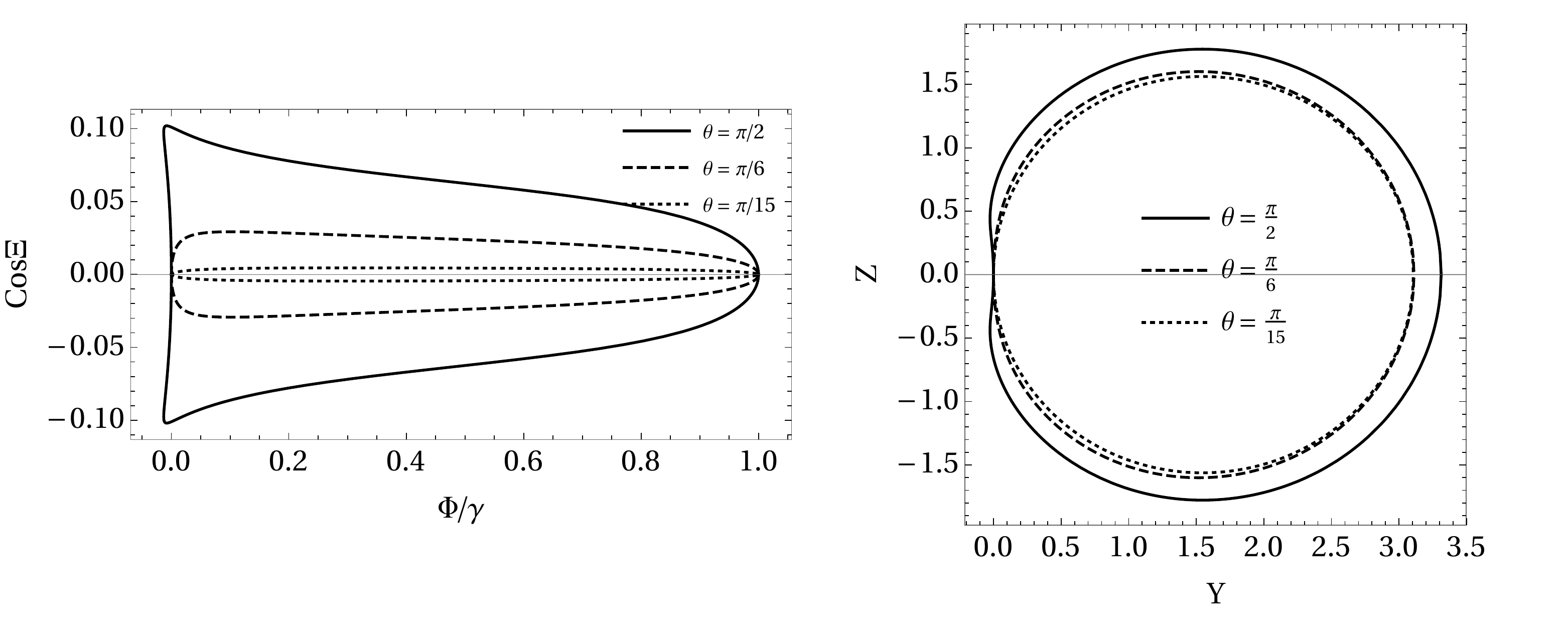}}
	\caption{Left panel: distortion parameter $\cos \Xi  $ as function of
		$\Phi / \gamma$ for different inclination $\theta$. Right panel: the appearance of the shadow in projection plane for different inclination. Here, we set observers' position $r = 4.1 M$.\label{F4a}}
\end{figure}

The 4-velocities of static observers take the form of
\begin{eqnarray}
  u_{\rm{stc}} & \equiv & \frac{1}{N} \partial_t ~.
\end{eqnarray}
In Figure~\ref{F3}, we present the angular diameter $\gamma$ as function of
inclination $\theta$. As shown in the right panel of Figure~\ref{F3},
the characteristic size of the shadow $\gamma$ decreases with the inclination $\theta$ for distant observers. In contrast, in the case that observers 
are close to the black hole, the $\gamma$ tends to increase with the
inclination, which is shown in the left panel of Figure~\ref{F3}. For the
latter case, it is partly because the shape of the shadow is less
distorted for observers at $r \rightarrow r_{c, +}$
{\cite{chang_revisiting_2020}}.
In Figure~\ref{F3A}, we show how to find the transition distance ($r_{\rm t}$) between the increase trend and decrease trend presented in Figure~\ref{F3}. Namely, the transition distance $r_{\rm t}$ is defined with a distance that the maximums of increase trend curves and maximums of decrease trend curves  are close in value. In right panel of Figure~\ref{F3A}, we present the transition distance $r_{\rm t}$ as function of spin of black hole. It shows that the transition distance decreases with spin of black holes. For spherical black holes, the $r_{\rm t}$ tends to spatial infinity, while for extreme rotating Kerr black hole, the $r_{\rm t}$ tends to the photon sphere. 
This conclusion also is showed in  Figure~{\ref{F4}} and \ref{F4a}, where we present the distortion parameter $\cos\Xi$ and appearance of
the shadow for selected inclination $\theta$. Besides, it is nothing surprising
that the shape of the shadow tends to be circular for the observers approaching the rotation axis of the black hole. 

\begin{figure}[h]
	\centering
  \includegraphics[width=.6\linewidth]{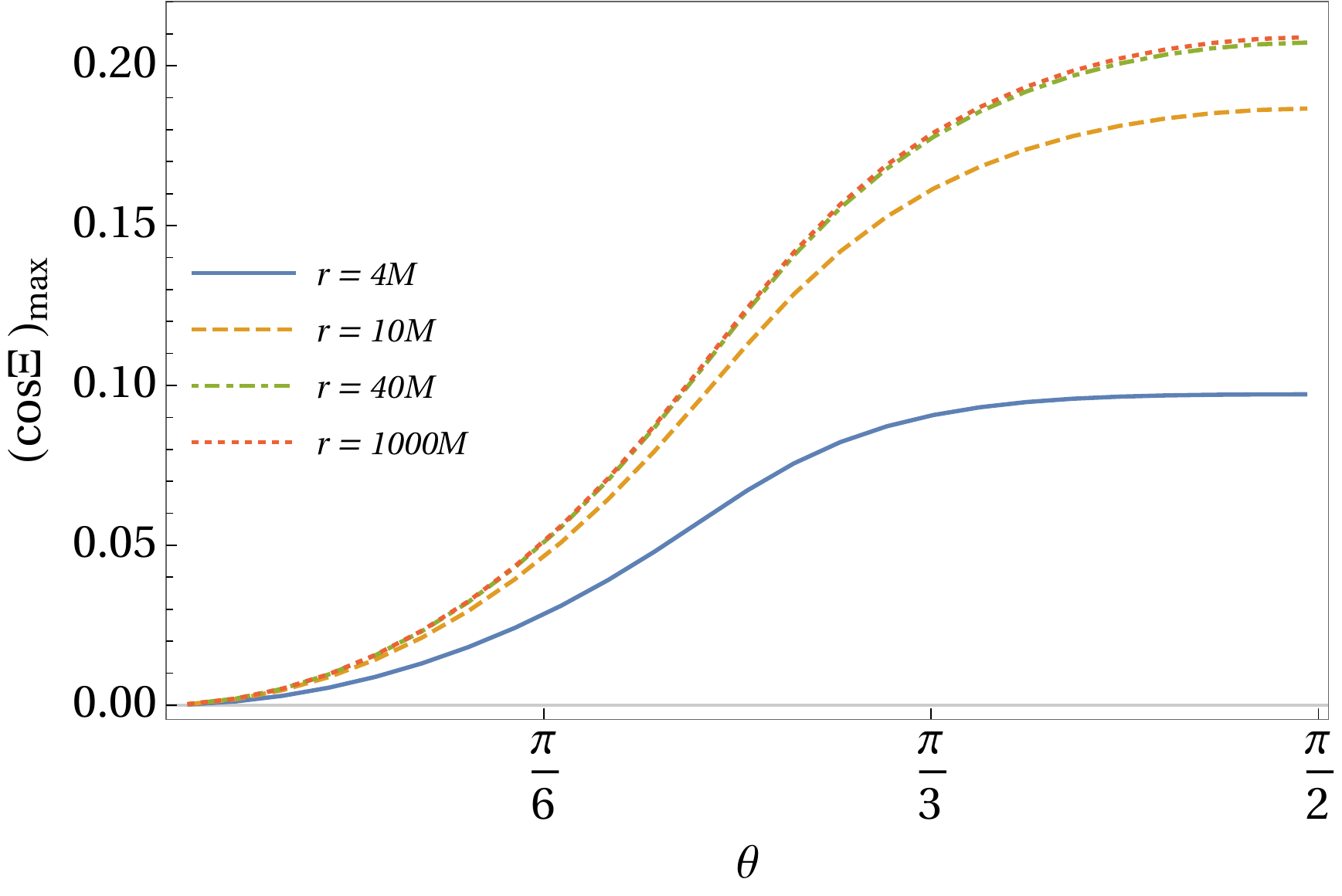}
  \caption{Maximum value of distortion parameter $\cos \Xi$ as function of
  inclination $\theta$.\label{F5}}
\end{figure}

In Figure~{\ref{F5}}, it shows the maximum value of the distortion
parameters as function of inclination for selected distance $r$. It is
consistent with the previous studies {\cite{chang_revisiting_2020,chang_does_2020}} that the distortion of the shadow get
larger with the distance of the observers from the black hole. The distortion parameter is
sensitive to the cases that observers are
close to the black hole, namely $r \rightarrow r_{c, +}$.

\subsection{Moving observers}

\begin{figure}[h]
	\centering
  \includegraphics[width=.4\linewidth]{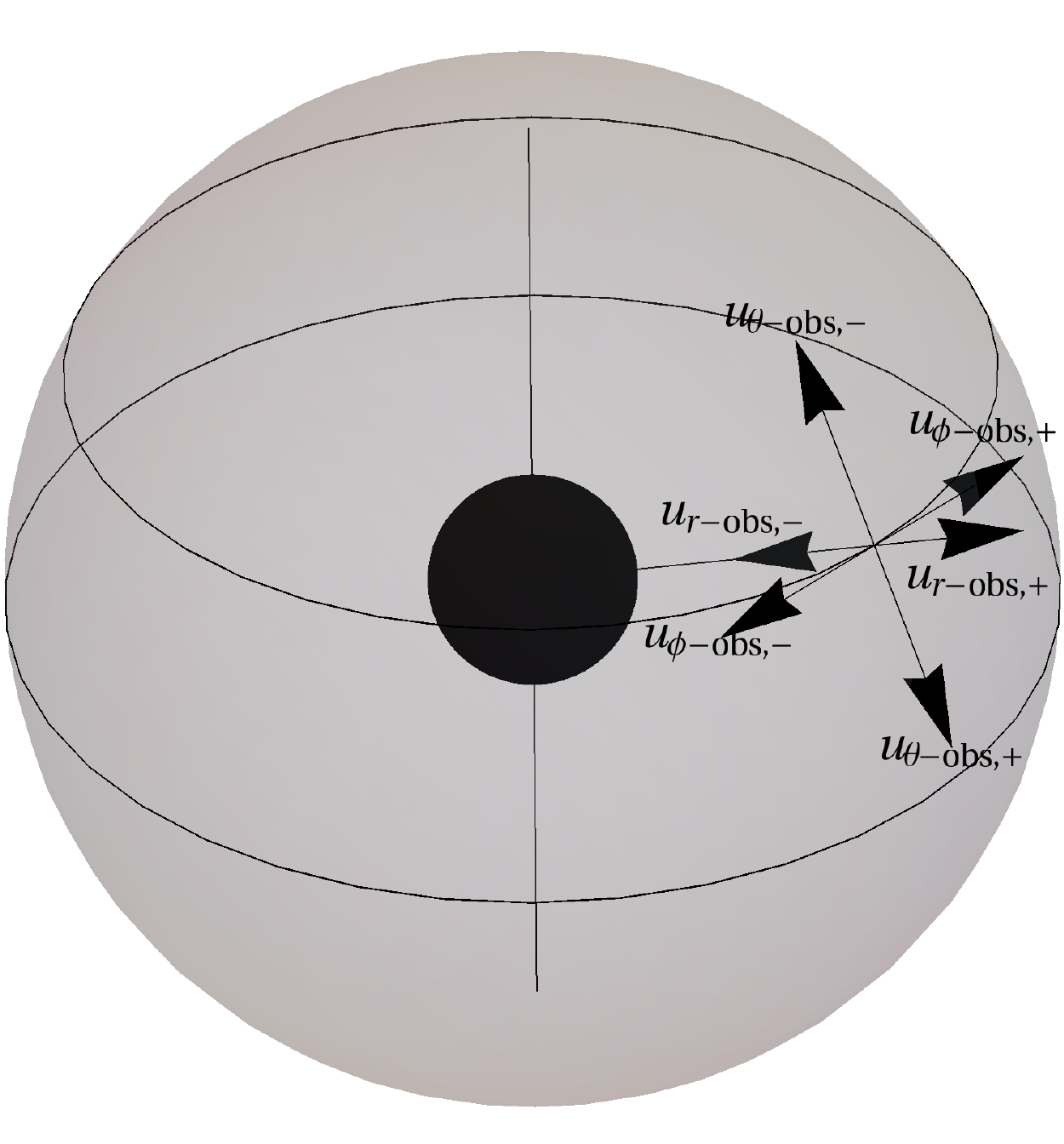}
  \caption{Schematic diagram for geodesic observers along the direction of increasing or decreasing
  	$r$, ${\phi}$ and ${\theta}$ Boyer-Lindquist coordinate. For
  illustration, the 4-velocities are labelled with $r$-obs, $\theta$-obs and
  $\phi$-obs.\label{F6}}
\end{figure}

In this part, we consider the cases that observers move along three representative
directions. Namely, they undergo geodesic motion along the direction of changing
$r$, ${\phi}$ and ${\theta}$ in Boyer-Lindquist
coordinate. The schematic diagram is shown in Figure~\ref{F6}. For
illustration, we would call these moving observers as $r$-observers,
$\theta$-observers and $\phi$-observers in the following.

The geodesic observers are the most natural observers in the present of
gravity. From Eq.~(\ref{A16}) with $\epsilon = - 1$, we have
4-velocities of the observers in the form of
\begin{subequations}
	\begin{eqnarray}
		u^t & = & \mathcal{E} \left( \frac{1}{N^2} - \frac{A}{G_{\phi}^2} (A
		+\mathcal{L}) \right) ~, \\
		u^r & = & \pm \mathcal{E} \frac{\sqrt{\Delta_r}}{B^2} \sqrt{f_1 (r ;
			\mathcal{L}) -\mathcal{K}- \left( \frac{B_r}{E} \right)^2} ~, \\
		u^{\theta} & = & \pm \mathcal{E} \frac{\sqrt{\Delta_{\theta}}}{B^2}
		\sqrt{\kappa + f_2 (\theta ; \mathcal{L}) - \left( \frac{B_{\theta}}{E}
			\right)^2} ~, \\
		u^{\phi} & = & \frac{\mathcal{E}}{G_{\phi}^2} (A +\mathcal{L}) ~ . 
	\end{eqnarray} \label{A29}
\end{subequations}
In order to distinguish above 4-velocities with that of light rays, we have
substituted the integral constant $E, \lambda$ and $\kappa$ in
Eqs.~(\ref{A16}) into $\mathcal{E}, \mathcal{L}$ and
$\mathcal{K}$ in Eqs.~(\ref{A29}), respectively. The trajectories of the geodesic observers are
not limited to circular orbits. It should be distinguished with the geodesic
observers in previous studies of black hole shadow {\cite{chang_does_2020}}.

For $r$-observers, they are instantaneously along the direction of $r$, i.e.
$u^{\theta} = u^{\phi} = 0$. Thus, we can rewrite the velocities by fixing the
constant constant $\mathcal{L}$ and $\mathcal{K}$,
\begin{eqnarray}
  u_{r \text{-} \rm{obs,\pm}} & = & \frac{\mathcal{E}}{N^2} \partial_t \pm
  \frac{1}{B} \sqrt{\Delta_r \left( \left( \frac{\mathcal{E}}{N} \right)^2 - 1
  \right)} \partial_r ~ .  \label{46}
\end{eqnarray}
Similar, for $\theta$-observers, we have
\begin{eqnarray}
  u_{\theta \text{-obs},\pm} & = & \frac{\mathcal{E}}{N^2} \partial_t \pm
  \frac{\sqrt{\Delta_{\theta}}}{B} \sqrt{\left( \frac{\mathcal{E}}{N}
  \right)^2 - 1} \partial_{\theta} ~ . 
\end{eqnarray}
And the 4-velocities of $\phi$-observers are
\begin{eqnarray}
  u_{\phi \text{-} \rm{obs,\pm}} & = & \left( \frac{\mathcal{E}}{N^2} \mp
  \frac{A}{G_{\phi}} \sqrt{\left( \frac{\mathcal{E}}{N} \right)^2 - 1} \right)
  \partial_t \pm \left( \frac{1}{G_{\phi}} \sqrt{\left( \frac{\mathcal{E}}{N}
  \right)^2 - 1} \right) \partial_{\phi} ~ .  \label{48}
\end{eqnarray}
The integral constant $\mathcal{E}$ is the only one left in Eq.~(\ref{46})--(\ref{48}). In the case of $\mathcal{E}= N$, above
velocities describe static observers. In order to compare the shadow
influenced by observers moving in different directions, we have to
introduce speed of these velocities. To be specific, we consider
the relative speed with respect to static observers
{\cite{felice_classical_2010}},
\begin{equation}
  \upsilon = \frac{| \gamma^{*} u |}{\sqrt{u_{\rm{stc}} \cdot u}} =
  \sqrt{1 - \left( \frac{N}{\mathcal{E}} \right)^2} ~,
\end{equation}
where the $\gamma^{*}$ is the projection operator with respect to the
static observers, i.e., $\gamma^{\mu}_{\nu} = u_{\rm{stc}}^{\mu}
u_{\rm{stc}, \nu} + \delta^{\mu}_{\nu}$. The second equal sign shows that
the 3-velocities are determined by the integral constant $\mathcal{E}$ and the
location of observers. The definition of the 3-speed might not be physical,
because the coordinate speed might not be measured directly in principle.
Here, we simply wish to make sense of different choice of $\mathcal{E}$. We
are not tended to present a well-defined 3-speed in curved space-time.

\subsubsection{ r-observers and $\phi$-observers}

\begin{figure}[h]
	\centering
  \includegraphics[width=\linewidth]{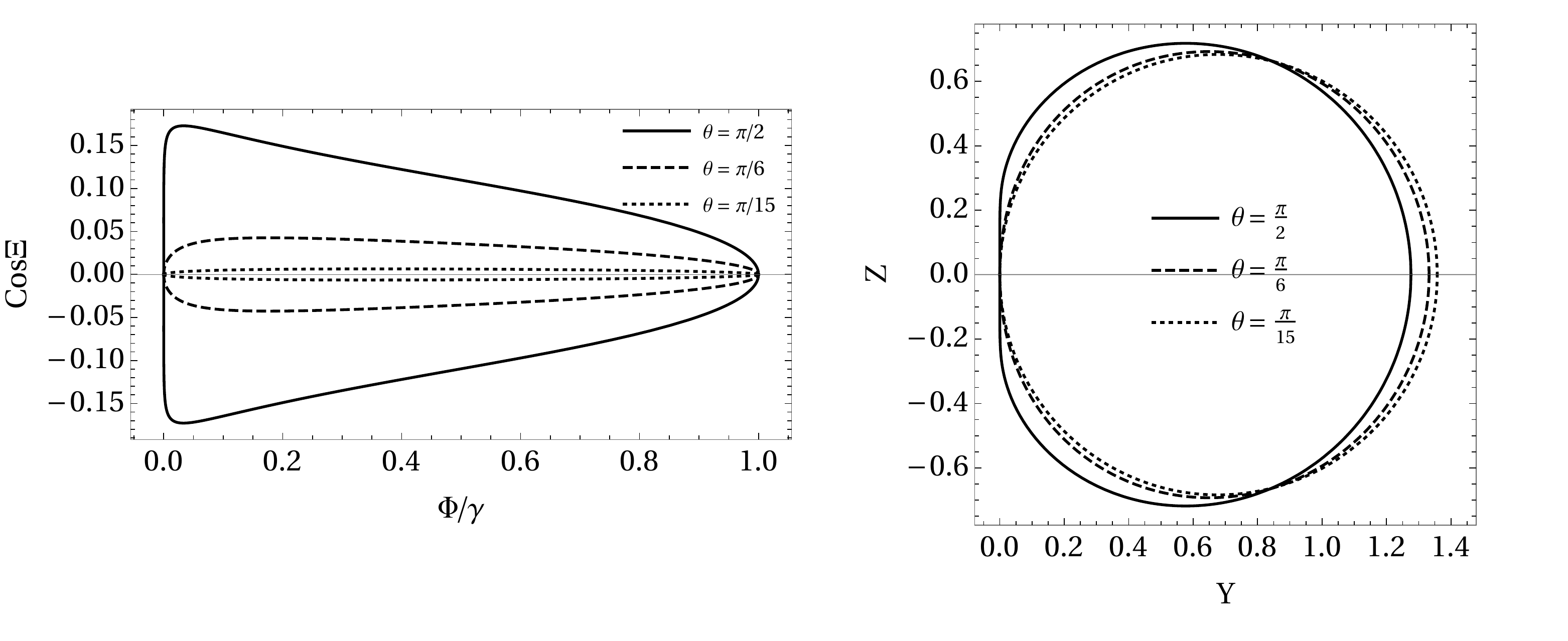}
  \caption{Left panel: distortion parameter $\cos \Xi  $ as function of
  $\Phi / \gamma$ for different inclination $\theta$. Right panel: the
  appearance of the shadow in projection plane for different inclination.
  Here, we set $r$-observers $u_{r \text{-} \rm{obs}, +}$ at $r = 10 M$ and $\upsilon = 0.3$.\label{F7}}
\end{figure}

\begin{figure}[h]
	\centering
  \includegraphics[width=\linewidth]{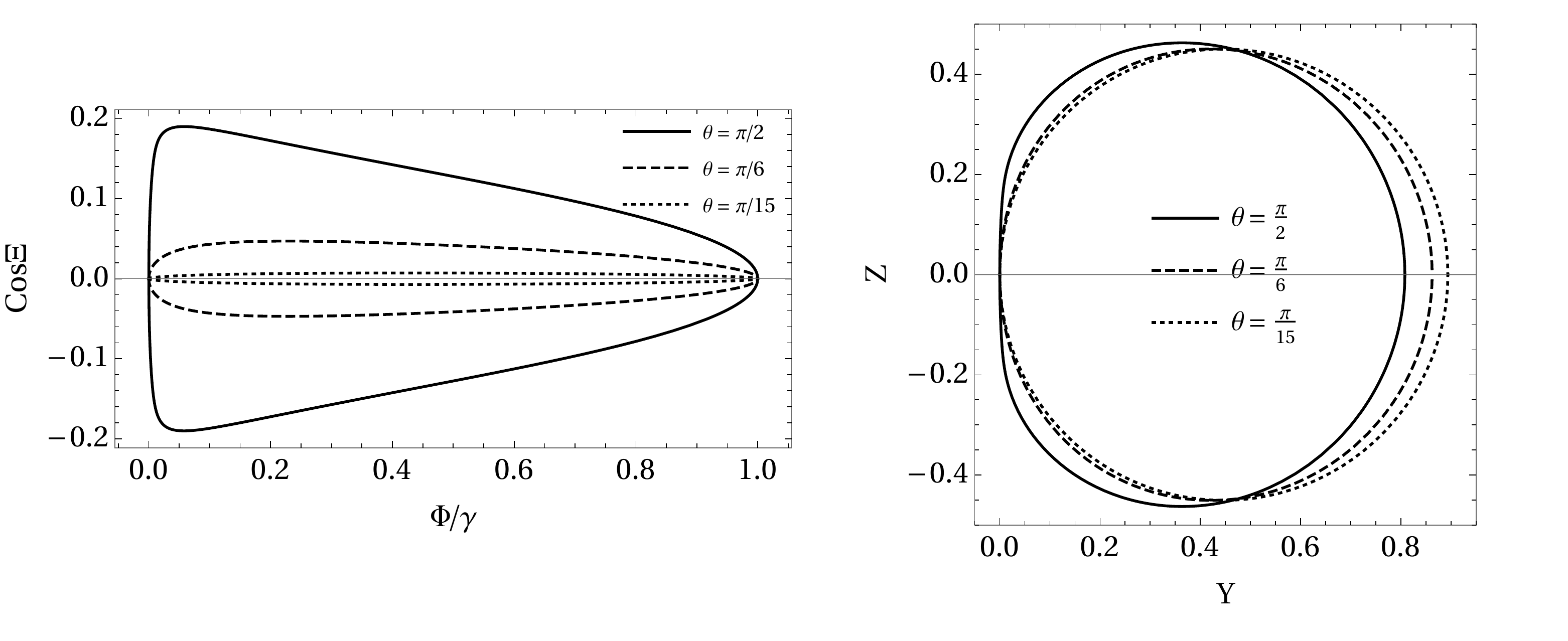}
  \caption{Left panel: distortion parameter $\cos \Xi  $ as function of
  $\Phi / \gamma$ for different inclination $\theta$. Right panel: the
  appearance of the shadow in projection plane for different inclination.
  Here, we set $\phi$-observers $u_{\phi \text{-} \rm{obs}, +}$ at $r = 10 M$ and $\upsilon = 0.3$. \label{F8}}
\end{figure}

\begin{figure}[h]
	\centering
  \includegraphics[width=\linewidth]{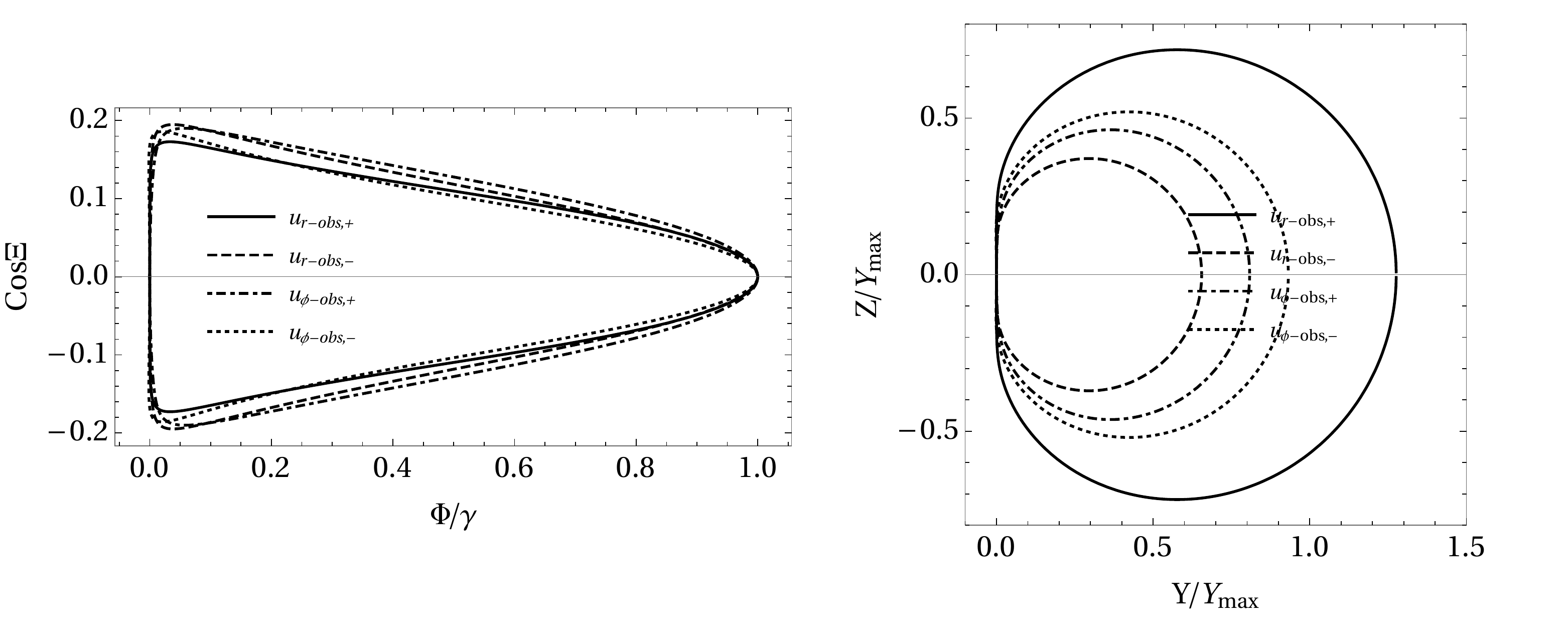}
  \caption{Left panel: distortion parameter $\cos \Xi  $ as function of
  $\Phi / \gamma$ for different observes. Right panel: the appearance of the
  shadow in projection plane for different inclination. Here, we set
  observers at $r = 10 M$, $\theta = \frac{\pi}{2}$ and $\upsilon=0.3$.\label{F9}}
\end{figure}

In Figure~\ref{F7}, we present the distortion parameter and appearance of
the shadow for selected inclinations $\theta$ in the view of
$r$-observers. And in Figure~\ref{F8}, we present the distortion parameter
and appearance of the shadow for selected inclination $\theta$ in the
view of $\phi$-observers. The shape of the shadow tends to be circular with
$\theta \rightarrow 0$. In Figure~\ref{F9}, we compare the distortion parameters
$\cos \Xi  $ and appearance of the shadows among $r$-observers and $\phi$-observes. Without the distortion
parameter, it seems difficult to tell the difference of the shape of the
shadow by eyes. As shown in the right panel of Figure~\ref{F9}, the size of the shadow  increases with $r$-components of the 4-velocities of $r$-observers, while it decreases with $\phi$-components of the 4-velocities of $\phi$-observers.


\begin{figure}[h]
	\centering
  \includegraphics[width=\linewidth]{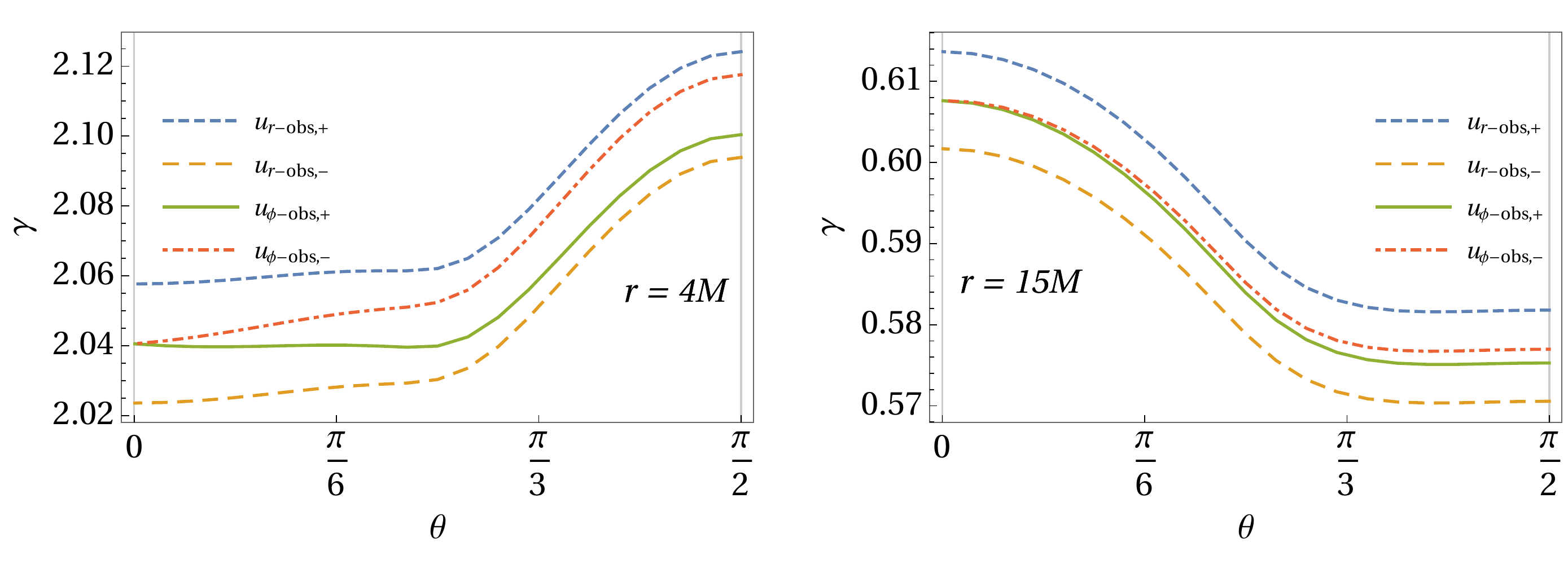}
  \caption{characteristic size $\gamma$ as function of inclination
  $\theta$ for different observers at the same speed $\upsilon=0.01$. The observes shown in the left panel are
  close to the black hole at $r = 4 M$, while the right panel describe the
  $\gamma$ with respect to more distant observers $r = 15 M$.  \label{F10}}
\end{figure}

In Figure~\ref{F10}, we show the characteristic size $\gamma$ as function
of inclination $\theta$ for selected observers at a fixing
distance. It is similar to that of static observers. For observers at different
distance, there is contrast tendency of characteristic size $\gamma$ changing with inclination $\theta$.

\subsubsection{$\theta$-observers}
\label{subsection}
\begin{figure}[h]
	\centering
  \includegraphics[width=\linewidth]{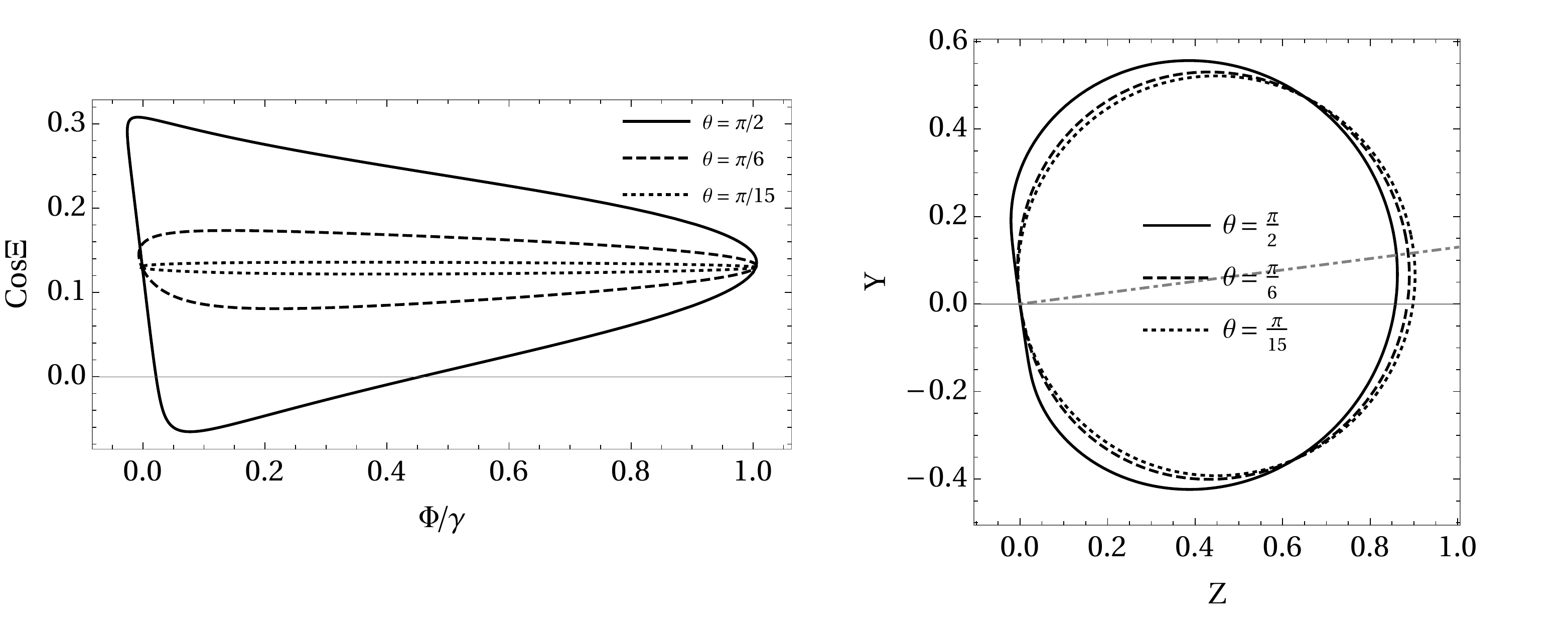}
  \caption{Left panel: distortion parameter $\cos \Xi  $ as function of
  $\Phi / \gamma$ for different inclination $\theta$. Right panel: the
  appearance of the shadow in projection plane for different inclination.
  Here, we set $\theta$-observers' $u_{\theta \text{-} \rm{obs}, +}$ at $r =
  10 M$, $\upsilon = 0.3$ and spin parameter $a=0.999M$. The angle $\Sigma$ is shown in dot-dashed line.\label{F11}}
\end{figure}

\begin{figure}[h]
	\centering
	\includegraphics[width=\linewidth]{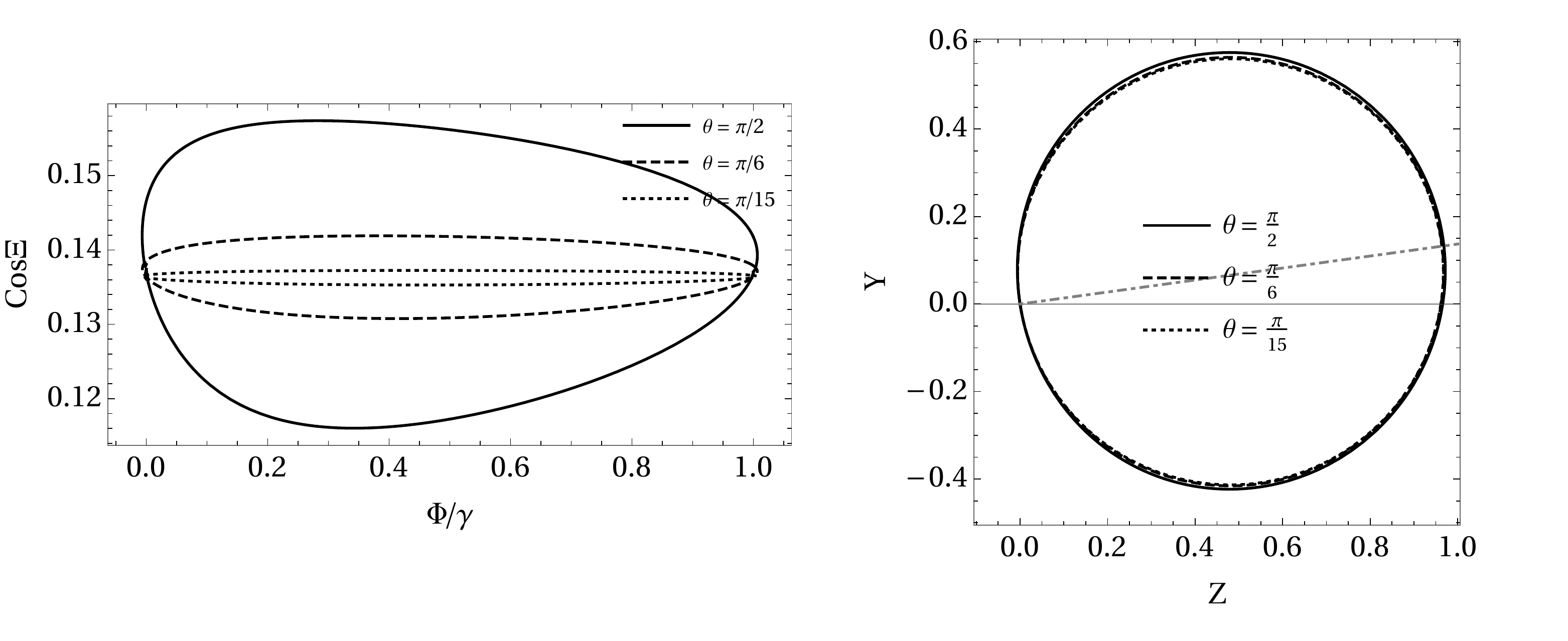}
	\caption{Left panel: distortion parameter $\cos \Xi  $ as function of
		$\Phi / \gamma$ for different inclination $\theta$. Right panel: the
		appearance of the shadow in projection plane for different inclination.
		Here, we set $\theta$-observers' $u_{\theta \text{-} \rm{obs}, +}$ at $r =
		10 M$, $\upsilon = 0.3$, and spin parameter $a=0.6M$. The angle $\Sigma$ is shown in dot-dashed line.\label{F11A}}
\end{figure}

For $r$-observers and $\phi$-observers, the influences of inclination is little different from that of the static observers. However, there would be
something new on the shadow in the view of the $\theta$-observers. In left
panel of Figure~\ref{F11}, we present the distortion parameter $\cos \Xi$
as function of $\Phi / \gamma$. It was expected that the shape of the shadow
tends to be circular for the observers at inclination $\theta= 0^{\circ}$. However, the distortion parameters $\Xi$ does not tends to $\pi / 2$ as
inclination $\theta \rightarrow 0^{\circ}$. In order to make it more
clear, we present the appearance of the shadow in the projection plane in right
panel Figure~\ref{F11}. We also present the plots in Figure~\ref{F11A} for a smaller spin parameter. The shapes of the shadow are similar to those in
the view of static observers. However, apparent difference is that the shadow seems to be rotated by a specific angle in the projection plane. 

Here, we should clarify whether it is caused by the way that we
calculate black hole shadow. In the case of observers located at rotation
plane of a rotating black hole, we have known it clearly
that the light rays $k$ and $w$  exactly propagates within the rotation
plane \cite{chang_revisiting_2020}. These reference light rays $k$ and $w$ should mark the fundamental plane
in observers' celestial sphere. One can confirm
that the point A and B in fundamental plane shown in
Figure~\ref{F2} always corresponds to the rotation plane of the black
hole. Therefore, the rotated image of the shadow shown in the right panel of Figure~\ref{F11} must be a physical result
caused by the motion of observers. The ways of the shadow getting distorted depend not only on the spin of the black hole, but also the velocities of observers.

\begin{figure}[h]
	\centering
  \includegraphics[width=.6\linewidth]{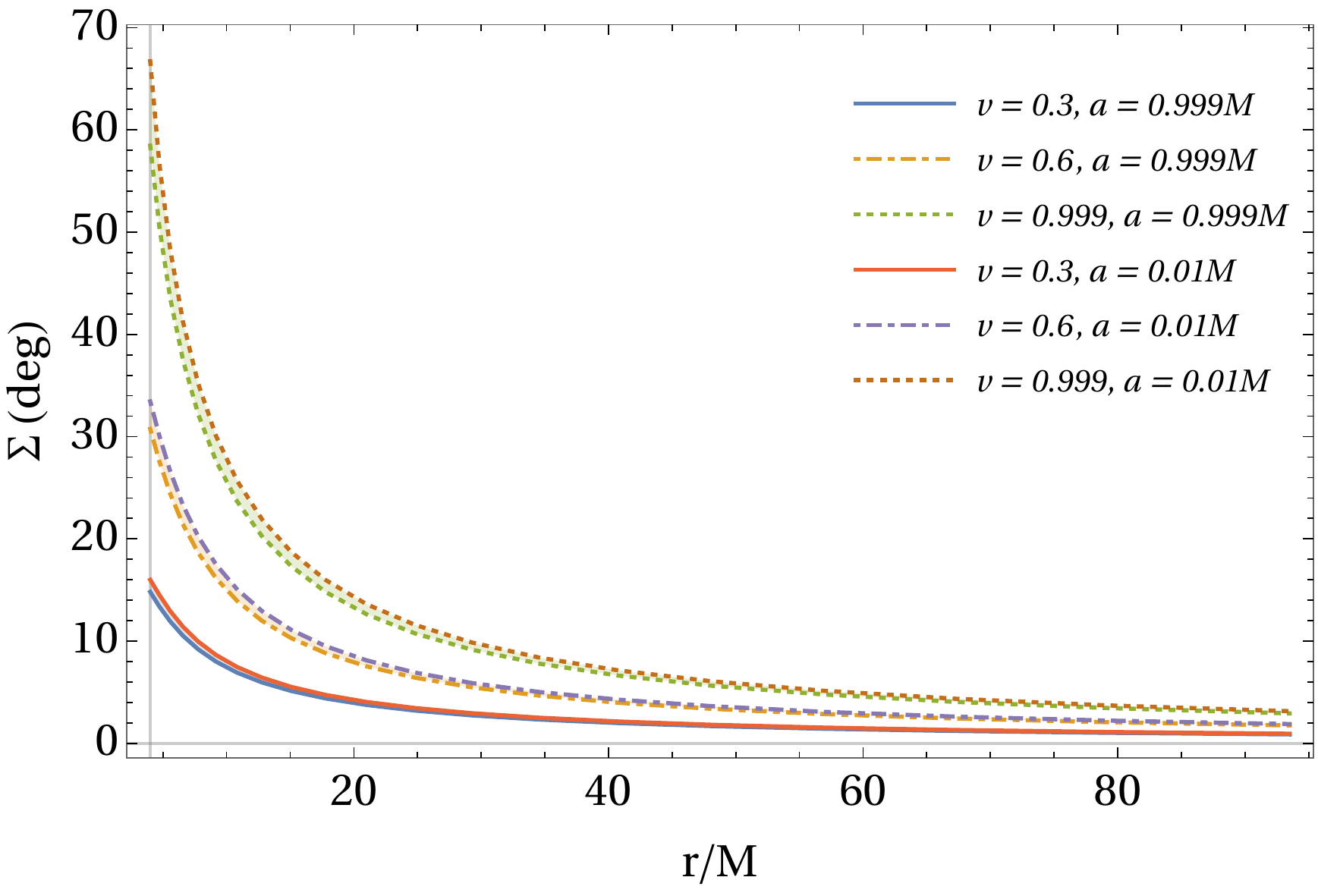}
  \caption{Rotation angle $\Sigma$ as function of coordinate $r$ of
  $\theta$-observers for selected speed $\upsilon$ and spin parameter $a$. \label{F12}}
\end{figure}

We quantify the rotation angle via the distortion parameter $\Xi$, which is
\begin{equation}
  \Sigma \equiv \frac{\pi}{2} - \Xi |_{\theta_{\rm{}} = 0} ~ \label{3.7}. 
\end{equation}
In the right panel of Figure~\ref{F11}, we also plot the angle $\Sigma$ in dotted-dashed line. It is basically derived from the geometric property of angle $\Xi$. In Figure~\ref{F12}. we present the rotation angle $\Sigma$ of image  as
function of the distance. The angle $\Sigma$ decreases with distance of
observers from the black hole. It indicates that the effect get more apparent for observers at finite distance. It could be understood that the observation for an observer in curved space-time is tended to be more non-trivial.
Besides, it is also interesting to consider whether the $\Sigma$ depends on the spin of a black hole. As shown in Figure~\ref{F12}, the spin might have not considerable influence on the $\Sigma$. Here, we also could not excluded the possibility that the deviations for different spin (etc. $\Sigma|_{a_1}-\Sigma|_{a_2}$) are originated from the trail definition of $\Sigma$ in Eq.~(\ref{3.7}), and is thus un-physical. 

\section{A formalism for velocity perturbation of distant observers}\label{IV}

As shown in Section~\ref{III}, motion of observers has influence on the
appearance of black hole shadow. Besides the geodesic motion around the centre
black hole, there are also influence from local orbital motions 
in a local gravitational potential, such as the EHT on the Earth orbits
the Sun in its local gravity potential. In this section, we will turn to this
local orbital motion. As these parts of
velocities of observers are usually very small ($\frac{\upsilon}{c} \simeq 10^{-
5}$ for the Earth's orbit), we consider the influence of local gravitational
potential via introducing expansion of velocities of observers, i.e.,
\begin{equation}
  u^{\mu} \rightarrow u^{(0), \mu} + \delta u^{\mu} ~ . \label{51}
\end{equation}
The $t$-component of the 4-velocity is determined by normalization
condition $u^{\mu} u_{\mu} = - 1$ and $u^{(0), \mu} u_{\mu}^{(0)} = - 1
+\mathcal{O} (\delta u)$. For illustration, we would call the changes of velocity $\delta u$ in Eq.~(\ref{51})  as local velocity.

By making use of the expansion of the velocities Eq.~(\ref{51}), we can
re-express the angular distances $(\alpha, \beta, \gamma)$ in
Eqs.~(\ref{E10}),
\begin{subequations}
	\begin{eqnarray}
		\hspace*{\fill} 1 - \cos \alpha & = & (1 - \cos \alpha^{(0)}) (1 + 2
		\delta_{\alpha}) ~, \\
		\hspace*{\fill} 1 - \cos \beta & = & (1 - \cos \beta^{(0)}) (1 + 2
		\delta_{\beta}) ~, \\
		\hspace*{\fill} 1 - \cos \gamma & = & (1 - \cos \gamma^{(0)}) (1 + 2
		\delta_{\gamma}) ~,  \label{54}
	\end{eqnarray}
\end{subequations}
where the $\alpha^{(0)}, \beta^{(0)}$ and $\gamma^{(0)}$ are the angular
distances with respect to the 4-velocity $u^{(0), \mu}$, and the relative
deviations of the angular distances $\delta_{\alpha}$, $\delta_{\beta}$ and
$\delta_{\gamma}$ are obtained to be
\begin{subequations}
	\begin{eqnarray}
		\delta_{\alpha} & = & - \frac{1}{2} \left( \frac{\delta u \cdot k}{u^{(0)}
			\cdot k} + \frac{\delta u \cdot l}{u^{(0)} \cdot l} \right) ~, 
		\label{55}\\
		\delta_{\beta} & = & - \frac{1}{2} \left( \frac{\delta u \cdot l}{u^{(0)}
			\cdot l} + \frac{\delta u \cdot w}{u^{(0)} \cdot w} \right) ~, \\
		\delta_{\gamma} & = & - \frac{1}{2} \left( \frac{\delta u \cdot k}{u^{(0)}
			\cdot k} + \frac{\delta u \cdot w}{u^{(0)} \cdot w} \right) ~ . 
		\label{57}
	\end{eqnarray}\label{A25}
\end{subequations}
For simplicity, above equations are expressed without indices. Here, the dot
product is defined as $u \cdot k \equiv g_{\mu \nu} u^{\mu} k^{\nu}$ ($\mu
= 0, 1, 2, 3$). As shown in Eqs.~(\ref{A25}), at the leading
order, the changes of the angular distances are proportional to local velocity  $\delta u^{\mu}$.

Using Eqs.~(\ref{37}) and (\ref{A25}), we can obtain the
relative deviation of distortion parameter in terms of local
velocity, namely,
\begin{eqnarray}
  1 - \cos \Xi & = & (1 - \cos \Xi^{(0)}) (1 + \delta_{\Xi}) ~, 
\end{eqnarray}
where
\begin{eqnarray}
  \delta_{\Xi} & = & \frac{\cos \Xi^{(0)}}{\cos \Xi^{(0)} - 1} \left( \frac{2
  (\cos \gamma - 1) \delta_{\gamma} - 2 (\cos \alpha - 1) \delta_{\alpha} - 2
  (\cos \beta - 1) \delta_{\beta}}{1 + \cos \gamma - \cos \alpha - \cos \beta}
  - \delta_{\alpha} - \delta_{\beta} \right) ~,  \label{59}
\end{eqnarray}
and the $\cos \Xi^{(0)}$ is distortion parameter with respect to 4-velocity
$u^{(0), \mu}$. From Eq.~(\ref{59}), one might find that the relative
deviation of the distortion parameter $\delta_{\Xi}$ is also proportional to
local velocity of an observer.

The formulae in Eqs.~(\ref{A25}) and (\ref{59})
are not limited to the cases of distant observers. In principle, local metric
perturbations should be introduced, in order to evaluate the local
velocity $\delta u$ in a curved space-time background. For the first step,
here we consider the local velocities of distant observers in an asymptotic
flatness space-time. In this case, the $\delta u$ can be easily obtained via a
local gravity potential in flat space-time background.

Besides, the appearance of the shadow can be also influenced by slightly
changes of local displacement in orbits. If considering this part of contributions,
one can obtain the formulae via expansion of $x^{\mu} \rightarrow x^{\mu} +
\delta x^{\mu}$. We would also discuss this contribution in the end of the section.

\subsection{Orbital observers}

For a more specific study based on the formalism introduced above, we turn to
consider observers orbiting around axes of $r$, $\theta$ and $\phi$. 
Namely, besides
the background speed $u^{(0)}$, there are additional velocities $\delta
\bm{u}$ from local orbits. For illustration, we will call
them $r$-rotation-observers, $\theta$-rotation-observers and
$\phi$-rotation-observers, respectively.

For an asymptotic flatness space-time and in the case of $r \gg M$, the metric is tended to be
Minkowski space-time. We derive the $\delta \bm{u}$ for distant observers
based on the Minkowski background. In this case, there is no ambiguous that
$\delta \bm{u}$ can describe physical velocities of the observers.

\begin{figure}[h]
	\centering
  \includegraphics[width=1\linewidth]{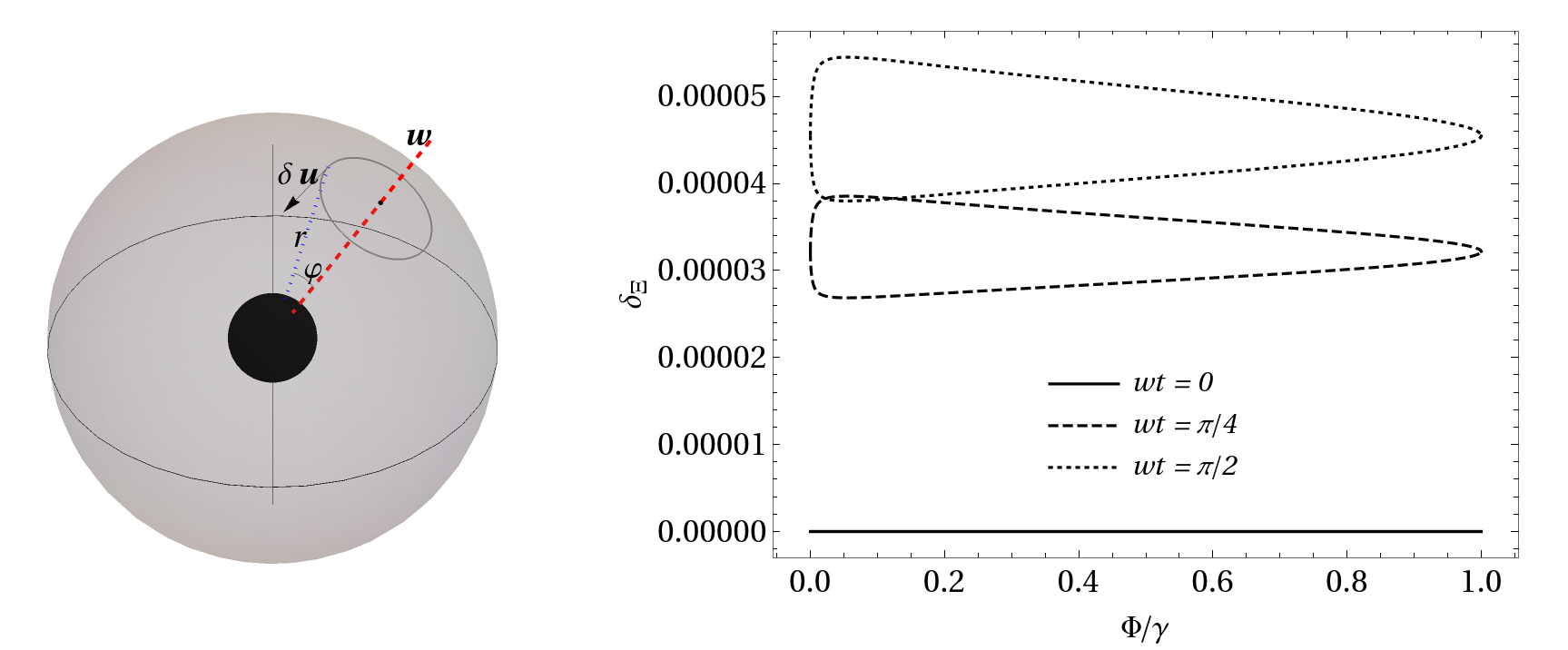}
  \caption{Left panel: schematic diagram of the orbit of the $r$-rotation-observers. Right panel: relative deviation of distortion parameter $\delta_{\Xi}$ as
  	function of $\Phi / \gamma$. Here, we set $r = 10^3 M$ and $\theta =
  	\frac{\pi}{3}$. And in order to present the influence of the $\delta
  	\bm{u}$ on the shadow, we set a quite large amplitude of the speed
  	$\upsilon = 10^{- 2} c$ for $r$-rotation-observers.
  \label{F13} \label{F14}}
\end{figure}

First, if considering rotation axis $\bm{w} = w \hat{\bm{r}}$, the
spatial part of local velocity $\delta \bm{u}$ takes the form of
\begin{eqnarray}
  \delta \bm{u}_{r \text{-rot-obs}} & = & - \frac{w \sin \varphi \sin (w
    t) \sin \theta}{\sqrt{1 - (\cos \theta \cos \varphi - \cos (w
    t) \sin \theta \sin \varphi)^2}} \partial_{\theta} \nonumber\\
  &  & + \frac{w \sin \varphi (\cos (w   t) \cos \varphi \sin \theta +
  \cos \theta \sin \varphi)}{1 - (\cos \theta \cos \varphi - \cos (w  
  t) \sin \theta \sin \varphi)^2} \partial_{\phi} ~, 
\end{eqnarray}
where the angle $\varphi$ is shown in left panel
Figure~\ref{F13}. In this case, the amplitude of  velocity $\delta
\bm{u}$ can be expressed as
\begin{equation}
  \upsilon \equiv \sqrt{\delta \bm{u}_{r \text{-rot-obs}} \cdot \delta
  \bm{u}_{r \text{-rot-obs}}} = w   r \sin \varphi ~ .
\end{equation}
In right panel Figure~\ref{F14}, we
present the relative deviation of distortion parameter $\delta_{\cos \Xi}$ as
function of $\Phi / \gamma$. For a fixing orbital
radius $d$ of detectors, i.e. $d \equiv r \sin \varphi$
(const.), the $\varphi$ tends to vanish as $r/M \rightarrow
\infty$. In the case of $w   t = 0$, we have $\delta \bm{u}_{r
\text{-rot-obs}} \propto \partial_{\phi}$, namely, there is no $\theta$-component of the local velocities. In contract, in the case of $w   t
\neq 0$, there exists  $\theta$-component of local velocities. It is consistent with results shown in Section~\ref{III} that the
parameter $\delta_{\Xi}$ deviated from zero is caused by observers' motion
along the direction of changing $\theta$.

\begin{figure}[h]
	\centering
  {\includegraphics[width=1\linewidth]{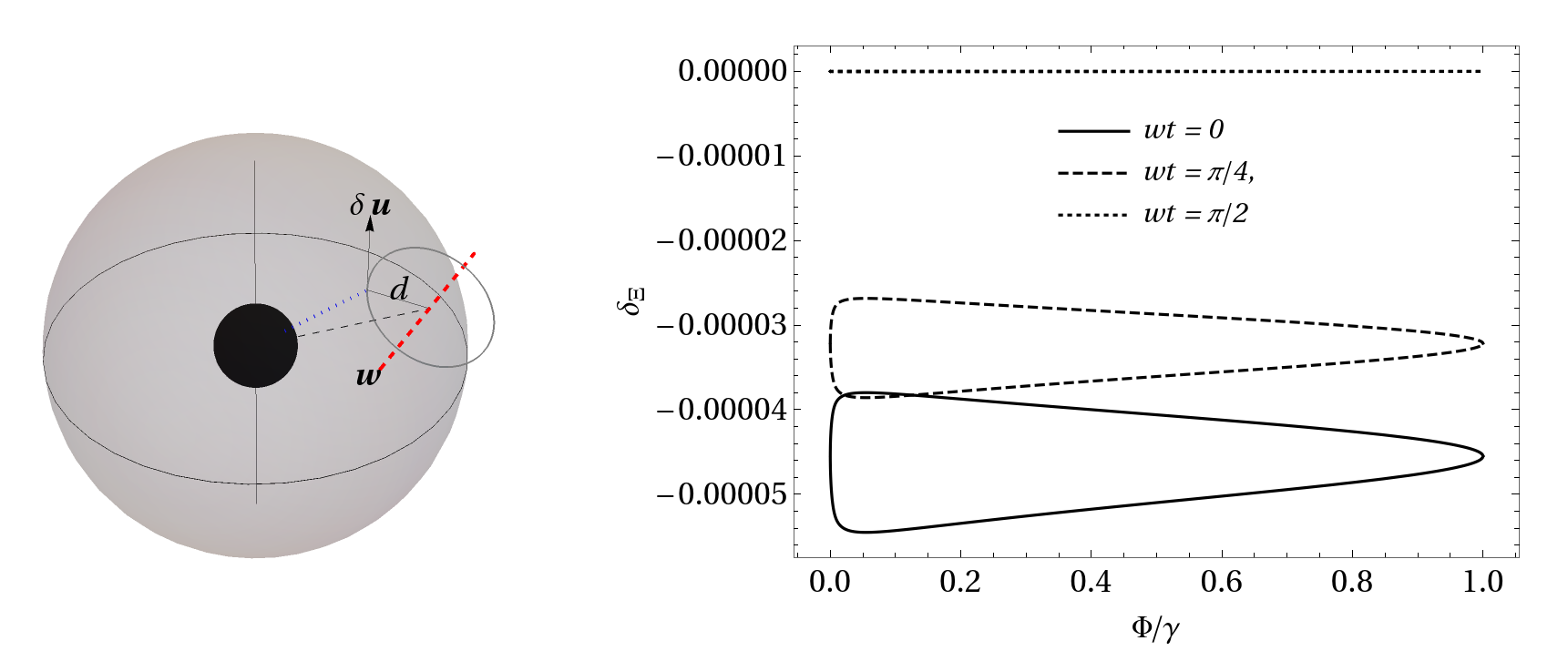}}
  \caption{Left panel: schematic diagram of the orbit of the $\phi$-rotation-observers. Right panel: relative deviation of distortion parameter $\delta_{\Xi}$ as
  	function of $\Phi / \gamma$. Here, we set $r = 10^3 M$, $\theta =
  	\frac{\pi}{3}$ and $\upsilon = 10^{- 2} c$ for $\phi$-rotation-observers.
  	\label{F16}   \label{F15}
}
\end{figure}

Second, for rotation axis $\bm{w} = w \left( \hat{\bm{z}}
\times \bm{r} \right)$ shown in Figure~\ref{F15}, we have spatial
part of local velocity $\delta \bm{u}$ in the form of
\begin{eqnarray}
  \delta \bm{u}_{\phi \text{-rot-obs}} & = & \frac{\upsilon \sin (w
    t)}{\sqrt{1 + \left( \frac{d}{r} \right)^2 + \frac{2 d}{r} \cos (w
    t)}} \partial_r + \frac{\upsilon}{r} \left( \frac{\frac{d}{r} +
  \cos (w   t)}{1 + \left( \frac{d}{r} \right)^2 + \frac{2 d}{r} \cos
  (w   t)} \right) \partial_{\theta} ~, 
\end{eqnarray}
where $d$ is orbital radius of a detector, and
\begin{eqnarray}
  \upsilon & \equiv & \sqrt{\delta \bm{u}_{\phi \text{-rot-obs}} \cdot
  \delta \bm{u}_{\phi \text{-rot-obs}}} = w   d ~ . 
\end{eqnarray}
In Figure~\ref{F16}, we present the relative deviation of distortion
parameter $\delta_{\cos \Xi}$ as function of $\Phi / \gamma$ for given $w
  t$. The tendency of the $\delta_{\Xi}$ for $\phi$-rotation-observers
is similar to that of $r$-rotation-observers. And in the case of of $w
  t = \pi / 2$, we have $\delta \bm{u}_{\phi \text{-rot-obs}}
\propto \partial_r$, and then the $\delta_{\Xi}$ tends to be zero.

\begin{figure}[h]
	\centering
  \includegraphics[width=\linewidth]{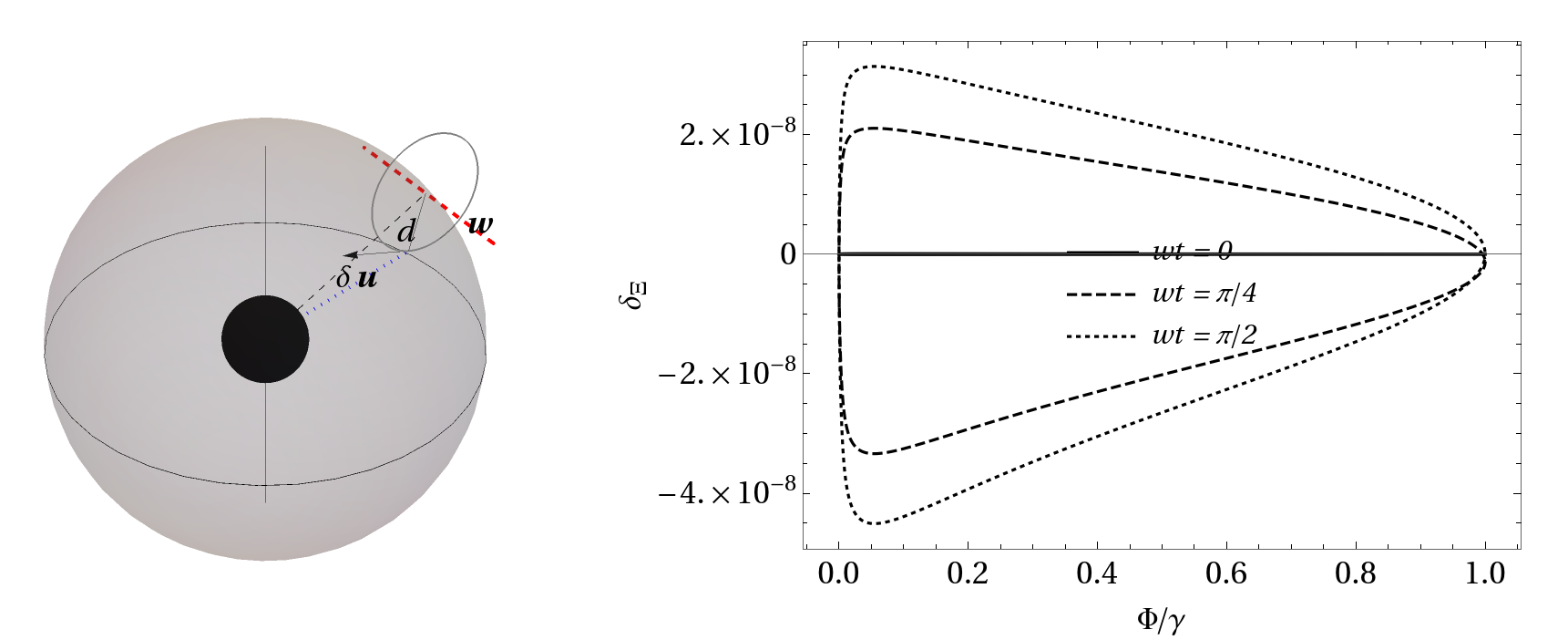}
  \caption{Left panel: schematic diagram of the orbit of the
  $\theta$-rotation-observers.\label{F17} Right panel: relative deviation of distortion parameter $\delta_{\Xi}$ as
  function of $\Phi / \gamma$. Here, we set $r = 10^3 M$, $\theta =
  \frac{\pi}{3}$ and $\upsilon = 10^{- 2} c$ for  $\theta$-rotation-observers.}
\end{figure}

Third, for rotation axis $\bm{w} = w \left( \left(
\hat{\bm{z}} \times \hat{\bm{r}} \right) \times
\hat{\bm{r}} \right)$ shown in Figure~\ref{F17}, spatial part
of local velocity $\delta \bm{u}$ is
\begin{eqnarray}
  \delta \bm{u}_{\theta \text{-rot-obs}} & = & \frac{d   r
    w \sin (w   t)}{\sqrt{d^2 + r^2 + 2 d   r \cos (w
    t)}} \partial_r \nonumber\\
  &  & + \frac{d^2 w \cos \theta \sin (w   t) (d + {r}\cos
    (w   t))}{(d^2 + r^2 + 2 d   {r}\cos (w  
  t)) \sqrt{2 d   r \cos (w   t) \sin \theta + r^2 \sin^2 \theta
  + d^2 (1 - \cos^2 (w   t) \cos^2 \theta)}} \partial_{\theta}
  \nonumber\\
  &  & + \frac{d   w \sin \theta (d + {r}   \cos(w t))}{r^2 \sin^2 \theta + 2 d   r   \cos (w   t) \sin^2
  \theta + d^2 (1 - \cos^2 (w   t) \cos^2 \theta)} \partial_{\phi} ~
  . 
\end{eqnarray}
In Figure~\ref{F17}, we present the relative deviation of distortion
parameter $\delta_{\cos \Xi}$ as function of $\Phi / \gamma$ for given $w
  t$ of $\delta \bm{u}_{\theta \text{-rot-obs}}$. In this case,
one can find the parameter $\delta_{\Xi}$ are much less than those of $\delta \bm{u}_{\phi
\text{-rot-obs}}$ and $\delta \bm{u}_{r \text{-rot-obs}}$.

Through the $\delta_{\Xi}$ for the three representative observers, it is found that the
$\theta$-component of the local velocities dominate the  $\delta_{\Xi}$. As shown in
Figure~\ref{F11},  one can no longer determine the rotation axis via the appearance of the shadow in practice. Fortunately, we might have alternative approach by making use of the quantity $\delta_{\Xi}$. Adjusting orbit of a
satellite for a minimum $\delta_{\Xi}$, then one can learn that the rotation axis of black hole and
the rotation axis of the orbit would turn to be coplanar in this adjusted orbit.
\label{A}

\subsection{Earth's orbit and Sun's Galactic orbit}

The EHT project has already aimed at the shadow of
Sgr A* in the centre of Milky way \cite{collaboration_first_2019-1}. In the future, lots of information about
the centre black hole would be collected. Therefore, it is necessary to consider
influence on the black hole shadow from physical point of view. As known that
the EHT are a set of telescopes located at the Earth. The Earth orbits the
Sun. And the Sun orbits the centre of the Milky way. There is local velocity
of the Earth with respect to the centre of Milky way around the order of magnitude
$10^{- 5} c$. It seems not unpractical to consider the influence of motion of
detectors or telescopes on the shadow of Sgr A*.

In section \ref{A}, we have shown that the characteristic size $\gamma$ and the
distortion parameter $\cos \Xi$ are affected by the local velocities in
Eqs.~(\ref{54}) and (\ref{59}). Besides, the Earth's orbit also present a
local displacement at the order of the orbital diameter. Thus, in this part, we
will study leading order effect of both the local displacements and local
velocities of detectors. In order to make these orbits more realistic, we
utilize the data of Sgr A* and solar system as shown in Table \ref{T1}. For
illustration, we simply assume Kerr black hole with spin parameter $a \simeq
1$ as a description of Sgr A*. In principle, it is not
difficult to extend the study on a more realistic black hole.

\setlength{\tabcolsep}{0.5mm}{
\begin{table}[h]
	  \caption{Data of Sgr A* and solar system \cite{reid_distance_1993,reid_proper_2004}\label{T1}}
  \begin{tabular}{lll}
  	\hline\hline
    Mass of Sgr A* ($M$) & $4 \times 10^6 M_{\odot}$ 
     & $M$\\
    Sun's distance  to galactic centre   ($R_{\odot}$) & $8 \rm{kpc}$  & $4.18 \times 10^{10}    \frac{G   M}{c^2}$ \\
    Galactic years ($T_{\odot}$) & $225 \rm{Myr}$ & $3.60 \times  10^{14} \frac{G   M}{c^3}$\\
     Earth's distance to sun ($d_{\oplus}$) &  $1 \rm{au}$ & $25.3 \frac{G   M}{c^2}$ \\
     Earth's  orbital angular velocities
    to sun  $(w_{\oplus})$ & $w_0 = \frac{2 \pi}{1 \rm{yr}}$, $ l =
    94.6^{\circ}$, $ b = 29.8^{\circ}$  & $w_0 = 3.93 \times
    10^{- 6} \frac{c^3}{G   M}$ \\
    \hline
  \end{tabular}
\end{table}
}

For two light rays from the photon sphere, angle $\psi$ between two incident
light rays in of observer's celestial sphere is
\begin{eqnarray}
  \cos \psi & = & \frac{p_1 \cdot p_2}{(u \cdot p_1) (u \cdot p_2)} + 1 ~, 
\end{eqnarray}
where $p_i \equiv p|_{r_c = r_{c, i}}$ and $u$ is 4-velocity of the Sun with respect to the central of the Milky way. The angular distance $\psi$ could be
$\alpha$, $\beta$ or $\gamma$. We expand the angle $\psi$ at $r / M
\rightarrow \infty$, namely
\begin{eqnarray}
  \cos \psi & = & 1 - \frac{\mathcal{C}_{\psi}^{(1)}}{r^2} -
  \frac{\mathcal{C}^{(2)}_{\psi}}{r^3} - \frac{\mathcal{C}_{\psi}^{(3)}}{r^4}
  +\mathcal{O} \left( \left(\frac{M}{r}\right)^4 \right) ~,  \label{66}
\end{eqnarray}
where
\begin{subequations}
\begin{eqnarray}
	\mathcal{C}_{\psi}^{(1)} & \equiv & \frac{\frac{\kappa_1 + \kappa_2}{2} -
		(\lambda_1 - a) (\lambda_2 - a) - \sqrt{\kappa_1 - (a - \lambda_1)^2}
		\sqrt{\kappa_2 - (a - \lambda_2)^2}}{\mathcal{E}^2} ~, \\
	\mathcal{C}_{\psi}^{(2)} & \equiv & \left( (\lambda_1 + \lambda_2) \sqrt{1 -
		\frac{1}{\mathcal{E}^2}} - 4 M \right) \mathcal{C}_{\psi} ~, \\
	\mathcal{C}_{\psi}^{(3)} & \equiv & \frac{1}{\mathcal{E}^2} \left(
	\frac{1}{8} (\kappa_1 - \kappa_2) (\kappa_1 - \kappa_2 + 4 a (\lambda_1 -
	\lambda_2)) \right) \nonumber\\
	&  & + \frac{\mathcal{C}_{\alpha}}{\sqrt{\mathcal{E}^2 - 1}} \Big(
	4\mathcal{E}M \left( a (\mathcal{E}^2 - 1) + M\mathcal{E}
	\sqrt{\mathcal{E}^2 - 1} \right) \nonumber\\
	&  & + (\mathcal{E}^2 - 1)^{\frac{3}{2}}
	(\lambda_1^2 + \lambda_2^2 + \lambda_1 \lambda_2) +\mathcal{E} (6 -
	5\mathcal{E}^2) M (\lambda_1 + \lambda_2) \Big) ~ . 
\end{eqnarray}
\end{subequations}
In above expressions, we have let $\lambda_i \equiv \lambda |_{r_c = r_{c,
i}}$, and $\kappa_i \equiv \kappa |_{r_c = r_{c, i}}$. The $\lambda$ and
$\kappa$ are functions of $r_c$ \ shown in Eqs.~(\ref{A-16}).
By making use of Eq.~(\ref{66}), we obtain the leading order of relative
deviation of the $\psi$ caused by local displacement $\Delta r$, i.e.
\begin{eqnarray}
  \delta_{\psi, \rm{disp}} \equiv \frac{\Delta \psi}{\psi} & = & -
  \frac{\Delta r}{r} +\mathcal{O} \left( \left(\frac{M}{r}\right)^2 \right) ~, 
\end{eqnarray}
This quantity have no relevance with coefficient
$\mathcal{C}^{(i)}_{\psi}$. Considering the shadow of Sgr A* observed by the
EHT, we can let $\Delta r \simeq 2 d_{\oplus}$ and $r \simeq r_{\odot}$.

Therefore, we can re-express the $\psi$ in the expansion both including the
contributions from local velocity and displacement,
\begin{eqnarray}
  \psi & = & \psi^{(0)} (1 + \delta_{\psi, \rm{velo}} + \delta_{\psi,
  \rm{disp}}) ~, 
\end{eqnarray}
where the expression of $\delta_{\psi, \rm{velo}}$ is similar to that of
$\delta_{\alpha}$, $\delta_{\beta}$ and $\delta_{\gamma}$, i.e.
\begin{equation}
  \delta_{\psi, \rm{velo}} = - \frac{1}{2} \left( \frac{\delta u \cdot
  p_1}{u^{(0)} \cdot p_2} + \frac{\delta u \cdot p_2}{u^{(0)} \cdot p_2}
  \right) ~, \label{72}
\end{equation}
In order to compare the relative deviation of local velocities $\delta_{\psi, \rm{velo}}$ with the $\delta_{\psi, \rm{disp}}$, we expand the
$\delta_{\psi, \rm{velo}}$ in Eq.~(\ref{72}) in terms of $r/M \rightarrow
\infty$,
\begin{eqnarray}
  \delta_{\psi, \rm{velo}} & = & \delta_{\psi, \rm{velo}}^{(0)} +
  \frac{1}{r} \delta^{(1)}_{\psi, \rm{velo}} +\mathcal{O} \left( \left(\frac{M}{r}\right)^2 \right) ~,  \label{73}
\end{eqnarray}
where
\begin{eqnarray}
  \delta^{(0)}_{\psi, \rm{velo}} & \equiv & \frac{1}{\mathcal{E}} (\delta
  u^r - \delta u^0) ~, \\
  \delta^{(1)}_{\psi, \rm{velo}} & \equiv & \frac{1}{2\mathcal{E}} \Bigg( 4
  M \delta u^0 + \upsilon_{\theta} \left( \sqrt{\kappa_1 - (\lambda_1 - a)^2}
  + \sqrt{\kappa_2 - (\lambda_2 - a)^2} \right) \nonumber \\
  & & + (\lambda_1 + \lambda_2)
  \left( \sqrt{1 - \frac{1}{\mathcal{E}^2}} (\delta u^r - \delta u^0) +
  \upsilon_{\phi} \right) \Bigg) ~ . 
\end{eqnarray}
As shown in Section~\ref{III}, it is found that $\delta u^t, \delta u^r
\simeq \mathcal{O} (1)$ and $\delta u^{\theta}$, $\delta u^{\phi} \simeq
\mathcal{O} \left( \frac{1}{r} \right)$ for distant observers. Therefore, in
Eq.~(\ref{73}), we have parametrized the local velocity as
\begin{eqnarray}
  \delta u^{\mu} & = & \left( \delta u^t, \delta u^r,
  \frac{\upsilon_{\theta}}{r}, \frac{\upsilon_{\phi}}{\rm{rsin} \theta}
  \right) ~ . 
\end{eqnarray}
And the Eq.~(\ref{73}) can be re-expressed as
\begin{eqnarray}
  \psi & = & \psi^{(0)} \left( 1 + \frac{\delta u^r - \delta u^0}{\mathcal{E}}
  + \frac{\delta^{(1)}_{\psi, \rm{velo}} - \Delta r}{r} +\mathcal{O} \left(
  \left(\frac{M}{r}\right)^2, \delta u \right) \right) ~ . 
\end{eqnarray}
The leading order term of deviation of $\psi$ is dominated by the local
velocities $\delta u^{\mu}$ and $\Delta r / r$. It has no relevance with
information of the black hole. Using the data in Table \ref{T1}, we
 compare different contributions of the deviation of $\psi$ in order of
magnitude, which is shown in Table \ref{T2}.

For the shape of the black hole shadow, we also expand the deviation of distortion
parameter $\cos \Xi$ in terms of $r / M \rightarrow \infty$ and $\delta u$,
namely,
\begin{eqnarray}
  1 - \cos \Xi & = & (1 - \cos \Xi^{(0)}) (1 + \delta_{\Xi} + \delta_{\Xi,
  \rm{disp}}) ~, 
\end{eqnarray}
where
\begin{eqnarray}
  \delta_{\Xi, \rm{disp}} & \equiv & - \frac{2 \Delta r}{r^3}
  \frac{1}{\mathcal{C}_{\alpha}^{(1)} +\mathcal{C}_{\beta}^{(1)}
  -\mathcal{C}_{\gamma}^{(1)} - 2 \sigma \sqrt{\mathcal{C}_{\alpha}^{(1)}
  \mathcal{C}_{\beta}^{(1)}}} \Bigg( \mathcal{C}_{\alpha}^{(3)}
  +\mathcal{C}_{\beta}^{(3)}  -\mathcal{C}_{\gamma}^{(3)} \nonumber\\
  & &-   (\mathcal{C}_{\alpha}^{(2)} +\mathcal{C}_{\beta}^{(2)}
  -\mathcal{C}_{\gamma}^{(2)}) \left(
  \frac{\mathcal{C}_{\alpha}^{(2)}}{2\mathcal{C}_{\alpha}^{(1)}} +
  \frac{\mathcal{C}_{\beta}^{(2)}}{2\mathcal{C}_{\beta}^{(1)}} \right) \nonumber\\
  & & +
  \left( \mathcal{C}_{\alpha}^{(1)} +\mathcal{C}_{\beta}^{(1)}
  -\mathcal{C}_{\gamma}^{(1)} \right) \left( \frac{3}{4} \left(
  \frac{\mathcal{C}_{\alpha}^{(2)}}{\mathcal{C}_{\alpha}^{(1)}} \right)^2 +
  \frac{3}{4} \left(
  \frac{\mathcal{C}_{\beta}^{(2)}}{\mathcal{C}_{\beta}^{(1)}} \right)^2 +
  \frac{\mathcal{C}_{\alpha}^{(2)}
  \mathcal{C}_{\beta}^{(2)}}{2\mathcal{C}_{\alpha}^{(1)}
  \mathcal{C}_{\beta}^{(1)}} -
  \frac{\mathcal{C}_{\alpha}^{(3)}}{\mathcal{C}_{\alpha}^{(1)}} -
  \frac{\mathcal{C}_{\beta}^{(3)}}{\mathcal{C}_{\beta}^{(1)}} \right) \Biggl)
  \nonumber\\
  & & +\mathcal{O} \left( \left(\frac{M}{r}\right)^4 \right) ~,
\end{eqnarray}
and we also expand expression of $\delta_{\Xi}$ in Eq.~(\ref{59}) at $r/M \rightarrow
\infty$
\begin{eqnarray}
  \delta_{\Xi} & = & \frac{1}{r} \left( \frac{- 2\mathcal{C}_{\gamma}^{(1)}
  \delta_{\gamma, \rm{velo}}^{(1)} + (\mathcal{C}_{\alpha}^{(1)}
  -\mathcal{C}_{\beta}^{(1)} +\mathcal{C}_{\gamma}^{(1)}) \delta_{\alpha,
  \rm{velo}}^{(1)} + (\mathcal{C}_{\beta}^{(1)} -\mathcal{C}_{\alpha}^{(1)}
  +\mathcal{C}_{\gamma}^{(1)}) \delta_{\beta, \rm{velo}}^{(1)}}{2
  \sqrt{\mathcal{C}_{\alpha}^{(1)} \mathcal{C}_{\beta}^{(1)}}} \right)
  \nonumber\\
  & & +\mathcal{O} \left( \left(\frac{M}{r}\right)^2 \right)~.
\end{eqnarray}
The $\delta_{\gamma, \rm{velo}}^{(1)}$, $\delta_{\alpha, \rm{velo}}^{(1)}$
and $\delta_{\beta, \rm{velo}}^{(1)}$ are relative deviations of angular
distances defined in Eqs.~(\ref{A25}). For distant observers, we
have $\delta_{\Xi} \simeq \delta u / r$ and $\delta_{\Xi, \rm{disp}} \simeq
\Delta r / r^3$. It suggests that the deviation of distortion parameter
$\delta_{\Xi} \gg \delta_{\Xi, \rm{disp}}$. Namely, local velocity $\delta
u$ has more significant influence on the shape of the shadow than the local
displacement $\Delta r$. In Table \ref{T2}, we present order of magnitude
estimation for the shape of the shadow of Sgr A*.

\setlength{\tabcolsep}{9mm}{
\begin{table}[h]
	  \caption{Order of magnitude of relative deviation of $\delta_{\gamma}$ and
		$\delta_{\Xi}  $ \label{T2} for the shadow of Sgr A*}
  \begin{tabular}{clll} 
  	\hline\hline   
     \multicolumn{2}{l}{Relative deviation}   & Estimators & Order of magnitude\\
    \hline 
    \multirow{2}{*}{$\delta_{\gamma}$} & $| \delta_{\gamma, \rm{velo}} |_{\max}$ & $\delta u
    / c$ & $10^{- 4}$\\    
    & $\delta_{\gamma, \rm{disp}}$ & $\frac{2 d_{\oplus}}{r}$ & $10^{-
    10}$\\
    \multirow{2}{*}{$\delta_{\Xi}$} & $| \delta_{\Xi} |_{\max}$ & $\frac{\delta u}{r} / \frac{G
      M}{c^3}$ & $10^{- 14}$\\
    & $| \delta_{\Xi, \rm{disp}} |_{\max}$ & $\frac{2
    d_{\oplus}}{r^3_{\odot}} / \left( \frac{c^2}{G   M} \right)^2$ &
    $10^{- 30}$\\
    \hline
  \end{tabular}
\end{table}
}

\begin{figure}[h]
	\centering
  \includegraphics[width=\linewidth]{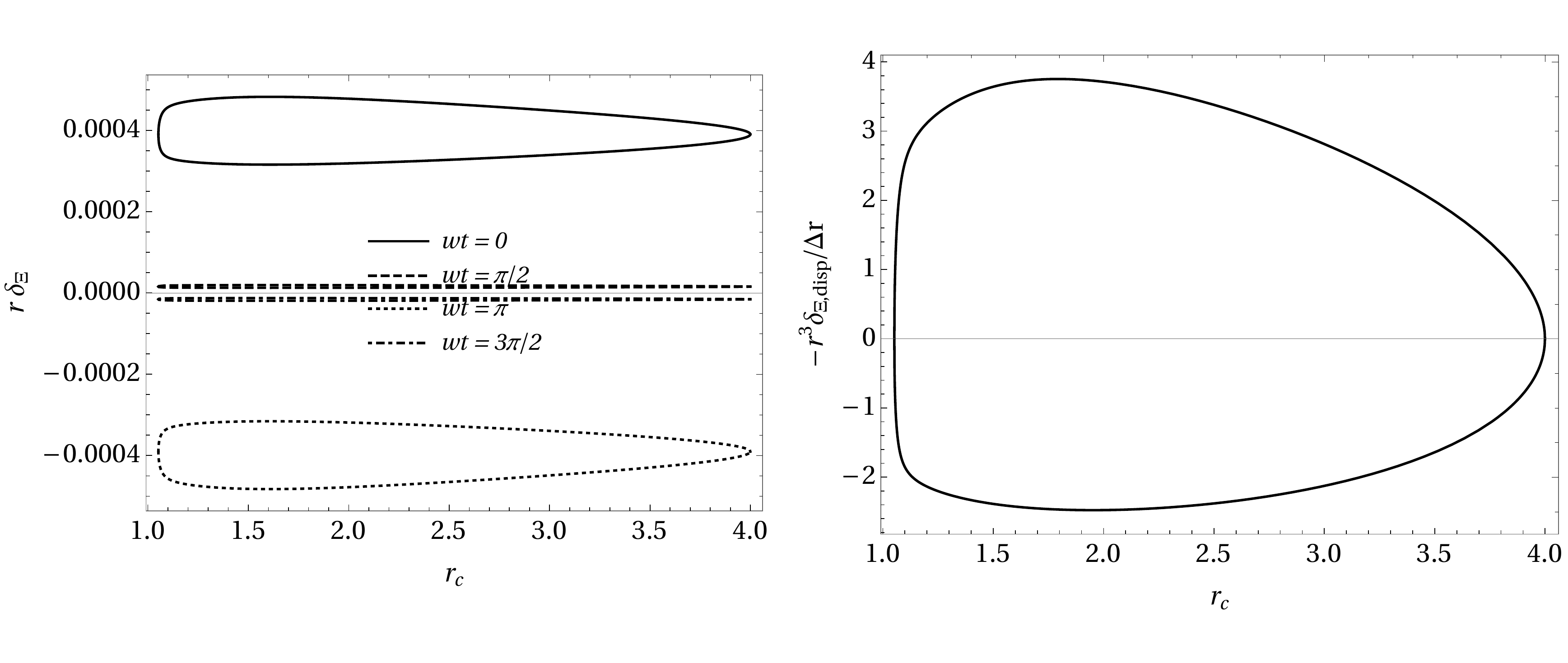}
  \caption{Left panel: deviation of distortion parameter of Earth's orbit
  $\delta_{\Xi}$ as function of $r_c$. Right panel: deviation of distortion
  parameter $\delta_{\Xi, \rm{disp}}$ of Earth's orbit as function of $r_c$. The 4-velocity describing
  the motion of the Earth is obtained by making use of the data of Sgr A* and
  solar system in Table~\ref{T1}\label{F19}}
\end{figure}

For the supermassive black hole Sgr A*, it is found that the local velocities of
observers dominate the effect on the size and shape of the black hole shadow.
Besides, we also present $\delta_{\Xi}$ and $\delta_{\Xi, \rm{disp}}$ as
functions of $r_c$ in Figure~\ref{F19}. It shows that the local velocities
and the local displacements have influence on the distortion parameter $\Xi$ in
a different way.

\section{Conclusions and discussions}\label{V}

In this paper, we investigated influence of Earth's orbit on the shadow of Sgr A*. We extended the previous studies \cite{chang_revisiting_2020,chang_does_2020} to arbitrary inclination, and observers at different velocities. It was partly motivated by the inclination of the Earth's orbit are not located at galactic plane \cite{reid_proper_2004}.  In this study, it was found that the appearance of the shadow would be rotated by a certain angle for an observer moving towards direction of changes of inclination $\theta$. 
It is beyond  pioneer's study \cite{bardeen_timelike_1973-1} that distorted shape of the shadow suggests spin direction of a rotating black hole.
In order to handle local orbits of observers with respect to the black hole Sgr A*, we presented a formalism for calculating the shadow in terms of the local velocity expansion. It showed that the influence of the orbital velocity of the Earth on the shadows is much larger than that of displacement in Earth's orbit. For the shadow of the Sgr A*, we obtained the deviation of characteristic size $\delta_{\gamma}$ of the shadow around $10^{-4}$. And the deviation of the distortion parameter $\delta_{\Xi}$ of the shadow is around $10^{-14}$.

It is found that rotation axis of a rotating black hole might not be extracted from its shadow. The ways of the shadow getting distorted depend not only on the spin of the black hole, but also the velocities of observers. This effect is shown to be less important for an observer at spatial infinity in an asymptotic flatness space-time. Perhaps, it might suggest that the observation for an observer in curved space-time is tended to be more non-trivial. 

In the present approach for calculating the black hole shadow, the reference light rays $k$ and $w$ are chosen to be the 4-velocities with vanished $\theta$-components for observers with $\theta =\pi/2$. It is still possible to define  reference light rays based on appearance of the shadow. Namely, one can let the reference rays locate the minimum diameter of the shadow in observers' celestial sphere. This scheme might be more practical for extracting information of a black hole from its shadow, although it could be rather artificial in formulae. 

Besides, as it shown in right panel of Figure~\ref{F11}, the shape of the shadow is seems to be unchanged. It might indicate that shape of the shadow from magnitude of spin is intrinsic prosperity of a rotating black hole. In succeeding studies, a more rigid study on this issue should be given.

\acknowledgments
 The authors wish to thank Prof.~Bin Chen for making mention of influence of spin of the Earth on the shadow of black hole.  This work has been funded by the National Nature Science Foundation of China  under grant No. 12075249 and 11690022.

\bibliography{citation}
\end{document}